\documentclass[%
reprint,
 amsmath,
 amssymb,
 aps,
 prl,
 superscriptaddress,
]{revtex4-2}
\usepackage{pdfpages}

\usepackage{graphicx}
\usepackage{dcolumn}
\usepackage{bm}
\newcommand{\ket}[1]{|{#1}\rangle}

\newcommand{\braket}[1]{\langle{#1}\rangle}
\usepackage[caption=false]{subfig}

\usepackage{mathtools}
\usepackage[T1]{fontenc}
\usepackage[utf8]{inputenc}
\usepackage{textcomp}
\usepackage[english]{babel}
\usepackage{lmodern}
\usepackage[babel]{microtype}

\DeclareUnicodeCharacter{2013}{--} 

\usepackage{xcolor}
\usepackage[colorlinks,
			linkcolor=black,
			citecolor=black,
			filecolor=black,
			urlcolor=blue!50!black,
			pdfusetitle]{hyperref}
\hypersetup{
	pdfauthor={Axel Fünfhaus, Titus Neupert, Thilo Kopp, Roser Valent\'i},
	pdfsubject={Solid-State Physics},
	pdfkeywords={topology, quantum Hall effect, interacting phases, time reversal symmetry breaking, fragile topology}
 }
\newcommand*\diff{\mathop{}\!\mathrm{d}}

\let\vec\boldsymbol

\makeatletter
\AtBeginDocument{\let\LS@rot\@undefined}
\makeatother

\begin{document}

\title{Many-body Euler topology}

\author{Axel Fünfhaus}
\email{fuenfhaus@itp.uni-frankfurt.de}
\affiliation{
 Institute of Theoretical Physics, Goethe University Frankfurt, Max-von-Laue-Straße 1, 60438 Frankfurt am Main, Germany
}
\author{Titus Neupert}
\affiliation{
 Department of Physics, University of Zurich, Winterthurerstrasse 190, 8057 Zurich, Switzerland
}
\author{Thilo Kopp}
\affiliation{
 Center for Electronic Correlations and Magnetism, Experimental Physics VI, Institute of Physics, University of Augsburg, 86135 Augsburg, Germany
}
\author{Roser Valent\'i}
\affiliation{
 Institute of Theoretical Physics, Goethe University Frankfurt, Max-von-Laue-Straße 1, 60438 Frankfurt am Main, Germany
}

\date{\today}

\begin{abstract}
Integer and fractional Chern insulators exhibit a nonzero quantized anomalous Hall conductivity due to a spontaneous breaking of time reversal symmetry. To identify nontrivial topology in their time-reversal symmetric many-body spectra, we introduce many-body Euler numbers as a counterpart to many-body Chern numbers. Exemplarily, we perform calculations in a topological Hubbard model that can realize Chern and fractional Chern insulating phases. Furthermore, we lay out a classification scheme to realize different topological phases in interacting systems using symmetry indicators in analogy to topological band theory.

\end{abstract}

\maketitle

\textit{Introduction.}---Chern and fractional Chern insulators~\cite{Neu2, Hal, Reg, She} are closely linked to the discovery of the integer and fractional quantum Hall effect~\cite{Kli, Tsu}. In contrast to quantum Hall systems, where time reversal invariance (TRI) is explicitly broken due to an external magnetic field, (fractional) Chern insulators can exhibit a nonzero ``anomalous'' Hall conductivity through the spontaneous breaking of TRI induced by interactions.  Conceived as a theoretical possibility initially, the experimental demonstration of such states of matter has been achieved in more than one material platform~\cite{Che, Wan, Wat, Zha, Zen} and remains the subject of ongoing theoretical inquiry~\cite{Cha}. In much of the existing literature, the breakdown of TRI is part of a computational framework from the outset, for instance by enforcing total spin conservation~\cite{Neu}, leading directly to magnetically ordered states or by using Hartree-Fock solutions that explicitly violate TRI~\cite{Li}. However, these approaches face conceptual difficulties. Enforcing total spin conservation, for example, fails for certain models of spinless fermions~\cite{Ulc, Zhu, Wu}  that exhibit an anomalous Hall effect.
Hartree-Fock theory, on the other hand, tends to overestimate the stability of symmetry-broken phases, particularly when modeling fractional Chern insulators, where solutions at integer fillings may become unstable at fractional fillings~\cite{Yu}. To avoid these intricacies, it is therefore desirable to directly trace the spontaneous breakdown of TRI within the time-reversal symmetric many-body spectrum.
 An issue arises, however, because the many-body Chern number, defined by using flux insertion~\cite{Niu}
 and commonly used to identify (fractional) Chern insulating phases in interacting systems~\cite{Haf, Var}, vanishes for time-reversal invariant spectra.

As is well known, in noninteracting topological band theory, symmetries can lead to new topological phases. A symmetry-protected integer invariant other than the Chern number is the Euler number. If a two-band subspace is characterized by a nonzero Euler number $e_2$, it describes two symmetry-related partners with opposite Chern numbers $\pm C$, with $|e_2| = |C|$. An example of their application in topological band theory is fragile topology~\cite{Ahn, Bou, Po, Ahn3}, where the label ``fragile'' implies that exactly two bands are needed for a well-defined Euler number.

In the present work, we want to use the concept of Euler numbers to identify the spontaneous breakdown of TRI in integer and fractional Chern insulators. As the ground state of an interacting system is not given by a product state of Bloch wave functions anymore, the standard topological band analysis is not applicable. Instead, we parameterize the Hamiltonian under periodic boundary conditions through the insertion of a flux that twists the boundary conditions, which allows defining topological invariants in finite-size clusters. We contrast both our approach and goal regarding the use of Euler numbers to previous efforts to identify signatures of single-particle fragile topology
in interacting systems, some of which also employed flux insertion~\cite{Son, Her, Els}.

In order to demonstrate the use of many-body Euler numbers we perform exact diagonalization (ED) calculations on a topological Hubbard model~\cite{Neu} which can realize time-reversal-symmetry-breaking ground states with Ising-ferromagnetic order. This model can be solved exactly in the limit of an infinite band gap and zero dispersion, making it a suitable toy model. After having calculated the Euler number with Wilson loop flows, we show that it is possible to have mathematically well-defined Euler numbers in fractional Chern insulators with topological ground state degeneracy using translation symmetry. In addition, we show how one can arrive at a classification scheme of topological phases protected by crystalline symmetries, focusing on rotation symmetry.

\textit{Chern insulator.}--- Our purpose is to define topological invariants with Euler numbers in interacting systems. Let us consider a finite two-dimensional (2D) cluster with periodic boundary conditions forming a torus. It is possible to insert fluxes through the noncontractible loops of the torus that minimally couple to itinerant particles via the Aharonov-Bohm effect. This can be realized as a twisted boundary condition (TBC) in the Hamiltonian, where particles hopping over the boundary of the cluster in $x$/$y$ direction obtain a phase factor $e^{i \theta_{x/y}}$, so that we can parameterize the many-body Hamiltonian as $\hat{H}(\vec{\theta})$~\cite{Niu, Sou, Sup}. Notably, the twist phase $\vec{\theta}$ --- which we can mathematically treat as a substitute for the Bloch wave vector $\vec{k}$, being periodic in $2\pi$ --- transforms like $\vec{k}$ under the action of symmetries. In particular, the time reversal operation $\hat{\mathcal{T}}$, like $\hat{C}_2$ (rotation by $\pi$), maps the Hamiltonian matrix to a gauge-equivalent matrix at $-\vec{\theta}$. However, in the case of $\hat{\mathcal{T}}$ --- being an antiunitary operation --- we also need to perform a complex conjugation. The Hamiltonian at arbitrary $\vec{\theta}$ is left invariant under the combined action of both symmetries, forming the antiunitary symmetry $\hat{A} = \hat{C}_2 \hat{\mathcal{T}}, \hat{A}^2 = +1$ (which in the following we assume not to depend on $\vec{\theta}$). The Hamiltonian matrix ${H}(\vec{\theta})$ is then equivalent to ${H}(\vec{\theta})^{*}$ up to a basis transformation. Furthermore, it can be shown that there exists a basis where ${H}(\vec{\theta})$ becomes real and hence has a set of real eigenstates, with $\hat{A}$ acting as complex conjugation on the basis states~\cite{Bou2}. This makes the definition of a new class of ``real'' topological invariants, namely Euler numbers, possible~\cite{Bou}, in contrast to Chern numbers that are nonzero only for complex states.

When a system is suspected to exhibit an anomalous Hall insulating phase, with interactions causing spontaneous breaking of time-reversal symmetry, we will expect to find two (quasi)degenerate states that are partners under time reversal. Upon flux insertion this can be generalized to $\ket{\Psi_{\pm C}(\vec{\theta})} \propto \hat{A} \ket{\Psi_{\mp C}(\vec{\theta})}$ with Hall conductivity $C$ and $-C$.
Due to finite-size splittings, the exact ground states of the Hamiltonian have to be written as $\ket{\Psi^{+}(\vec{\theta})} = (\ket{\Psi_{+C}(\vec{\theta})} + \ket{\Psi_{-C}(\vec{\theta})})/\sqrt{2}$ and $\ket{\Psi^{-}(\vec{\theta})} = i (\ket{\Psi_{+C}(\vec{\theta})} - \ket{\Psi_{-C}(\vec{\theta})})/\sqrt{2}$ (where we assume $\ket{\Psi^{\pm}(\vec{\theta})}$ to have only real vector entries). This allows us to define the Euler number~\cite{Gua, Nak}
\begin{equation}
    e_2 = \frac{1}{2\pi} \int_{\text{T}^2_{\text{TBC}}} \diff^2 \vec{\theta} \mathcal{F}_{+-}(\vec{\theta}),
\end{equation}
with
\begin{equation}
    \mathcal{F}_{+-}(\vec{\theta}) = \left. \vec{\nabla}_{\vec{\theta}} \times \braket{\Psi^{+}(\vec{\theta})|\vec{\nabla}_{\vec{\theta}}|\Psi^{-}(\vec{\theta})} \right|_{z}
\end{equation}
and $\text{T}^2_{\text{TBC}}$ is the torus of TBCs. It is easy to show (see Ref.~\cite{Sup}, which includes Refs.~\cite{Kob, Lin, Nas, Bra2, Bor, Bro, Fig, Pan, Fun2, Bou3, Gir, Pau, Wat3}) that the ``Euler curvature'' $\mathcal{F}_{+-}(\vec{\theta})$ is related to the Berry curvature
\begin{equation}
    F^{\pm}(\vec{\theta}) = \vec{\nabla}_{\vec{\theta}} \times i  \braket{\Psi_{\pm C}(\vec{\theta})|\vec{\nabla}_{\vec{\theta}}|\Psi_{\pm C}(\vec{\theta})}
\end{equation}
according to $\mathcal{F}_{+-}(\vec{\theta}) = (F^{+}(\vec{\theta}) - F^{-}(\vec{\theta}))/2$,  hence
\begin{equation}
    |e_2| = |C|,
\end{equation}
so the Euler number yields the Hall conductivity. It should be noted that the sign of the Euler number is not a topological invariant and corresponds to a gauge freedom in choosing the sign of the real eigenstates $\ket{\Psi^{\pm}}$~\cite{Bou}. Similarly to Chern numbers, it is possible to show that $|e_2|$ is determined by the Wilson loop operator $W[\mathcal{C}]$~\cite{Bou, Sup, Bou2, Ahn2}, defined on a closed, noncontractible loop $\mathcal{C}$. For Hamiltonians commuting with $\hat{A}$, its eigenvalues are either real or given by pairs $e^{\pm i \varphi[\mathcal{C}]}$. The number of loops of the phases $\varphi[\mathcal{C}]$ is equal to $|e_2|$, and easier to evaluate numerically then calculating $\mathcal{F}_{+-}(\vec{\theta})$. For calculations, we use the periodic gauge, which asigns a phase factor of $e^{-i {\theta_x}/{N_x}}$ and $e^{-i {\theta_y}/{N_y}}$ whenever a particle hops in $x$ and respectively $y$ direction, to realize twisted boundary conditions~\cite{Wat2}. We define the Wilson loop operator
\begin{equation}
    W(\theta_y) = \lim_{N \to \infty} \mathcal{F}_{0}(\theta_y) \dots \mathcal{F}_{N-1}(\theta_y),
\end{equation}
where
\begin{equation}
    \mathcal{F}_{0}(\theta_y)_{m,n} = \braket{\Psi_{m}(\theta_x^{j}, \theta_y)|\Psi_{n}(\theta_x^{j+1}, \theta_y)}, \quad \theta_x^{j} = 2\pi j / N,
\end{equation}
and for gauge invariance 
\begin{equation}
    \ket{\Psi_{\alpha}(2\pi, \theta_y)} := \hat{U}_{x} \ket{\Psi_{\alpha}(0, \theta_y)}.
\end{equation}
Here, $\alpha$ runs over all ground states and $\hat{U}_{x}$ is a large gauge transformation
\begin{equation}
    \hat{U}_{x} = \exp \left( - \frac{2\pi i}{N_x} \displaystyle\sum_{x,y} \displaystyle\sum_{\sigma} x c_{x,y,\sigma}^{\dagger} c_{x,y, \sigma} \right),
\end{equation}
$\hat{H}(\theta_x + 2\pi, \theta_y) = \hat{U}_{x} \hat{H}(\theta_x, \theta_y) \hat{U}_{x}^{\dagger}$. We choose to calculate the Euler number via Wilson loop flows in the following, but we note that determining the Hall conductivity via Euler curvature could be particularly useful for costly calculations with few accessible data points~\cite{Kud}.

To show how Wilson loop flows can trace the ground state topology,
we consider a variant of a topological Hubbard model on the square lattice defined in Ref.~\onlinecite{Neu}. The hopping term is a TRI Hofstadter Hamiltonian~\cite{Hof} with $\pi$ flux per plaquette and a spin-dependent Peierls' phase
\begin{widetext}
\begin{equation}\label{Eq:kinetic_hamiltonian}
\begin{aligned}[b]
    \hat{H}_{\text{kin}} = \displaystyle\sum_{x, y = 1}^{N_x, N_y} &\left[ \displaystyle\sum_{\sigma, \sigma' = \uparrow, \downarrow} -t \left( c_{x+1, y, \sigma}^{\dagger} \exp(-2\pi i \gamma \sigma_x)_{\sigma, \sigma'} c_{x,y, \sigma'}  + (-1)^{x} c_{x, y+1, \sigma}^{\dagger} \exp(-2\pi i \gamma \sigma_y)_{\sigma, \sigma'} c_{x,y, \sigma'} \right) \right.\\
    & \quad  \displaystyle\sum_{\sigma = \uparrow, \downarrow} - t' (-1)^x \left( e^{-i \frac{\pi}{2} \sigma} c_{x+1, y+1, \sigma}^{\dagger} c_{x,y, \sigma}  \left. + e^{i \frac{\pi}{2} \sigma} c_{x+1, y-1, \sigma}^{\dagger} c_{x,y, \sigma} \right) \right] + \text{h.c.}
\end{aligned}
\end{equation}
\end{widetext}
The model has a unit cell comprising two sites in $x$ direction. For $\gamma = 0$, the hopping term conserves the total spin in $z$ direction $S^{z}$ and has two gapped bands containing spin-up and spin-down particles. The Chern number of the spin-up/spin-down particles in the lower band is then $+1$ and $-1$, respectively. To break TRI spontaneously, we add the Hubbard term
\begin{equation}
    \hat{H}_{U} = U \displaystyle\sum_{x,y} \hat{n}_{x,y,\uparrow} \hat{n}_{x,y, \downarrow}.
\end{equation}
At half filling (one particle per unit cell), if $U$ is much larger than the band width but much smaller than the band gap, for $\gamma = 0$ the Hamiltonian will  realize two spin-polarized ground states to minimize the energy cost associated with the Hubbard term~\cite{Neu}. In the limit of no band dispersion and an infinitely large band gap, which we can realize as
\begin{equation}\label{Eq:projected_Hamiltonian}
    \hat{H} = \hat{P} \hat{H}_{U} \hat{P},
\end{equation}
where $\hat{P}$ projects the states to the lower band of Eq.~\eqref{Eq:kinetic_hamiltonian}, the ground states are exactly given as
\begin{equation}
    \ket{\Psi_{\sigma}} = \displaystyle\prod_{\vec{k} \in \text{BZ}} \bar{c}^{\dagger}_{-, \vec{k}, \sigma} \ket{0},
\end{equation}
where $\bar{c}^{\dagger}_{-, \vec{k}, \sigma}$ creates particles in the lower band with $s^{z} = \sigma$ and momentum $\vec{k}$ in the Brillouin (BZ). $\ket{0}$ is the vacuum state. Flipping a single spin would lead to a finite energy penalty due to the Hubbard term, as the $\bar{c}^{\dagger}_{-, \vec{k}, \sigma}$ operators will have finite weight in most, if not all, lattice sites~\cite{Neu}. Since both these states occupy all spin-up/spin-down particles in the lower band, they realize an anomalous, TRI-breaking quantum Hall effect with opposite Chern number~\cite{Sup}.

\begin{figure}[t] 
\vspace{-0.3cm}
    \centering
    \begin{minipage}[b]{0.48\columnwidth} 
        \includegraphics[width=\linewidth]{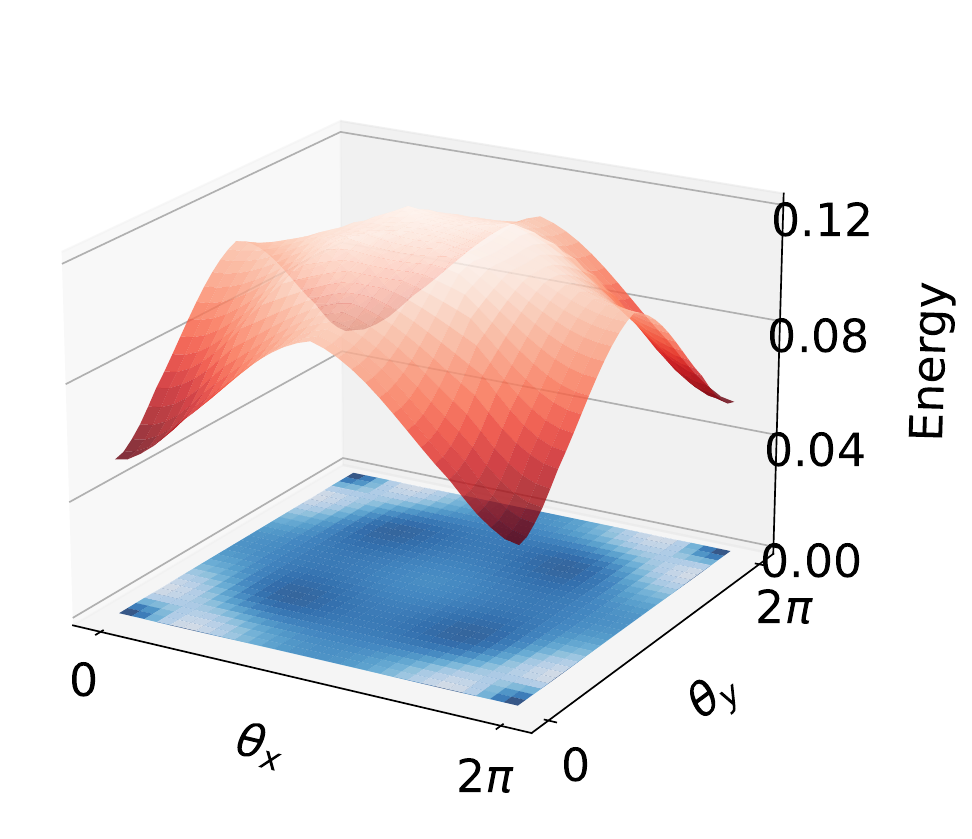}
        \centering
    \includegraphics[width=\linewidth]{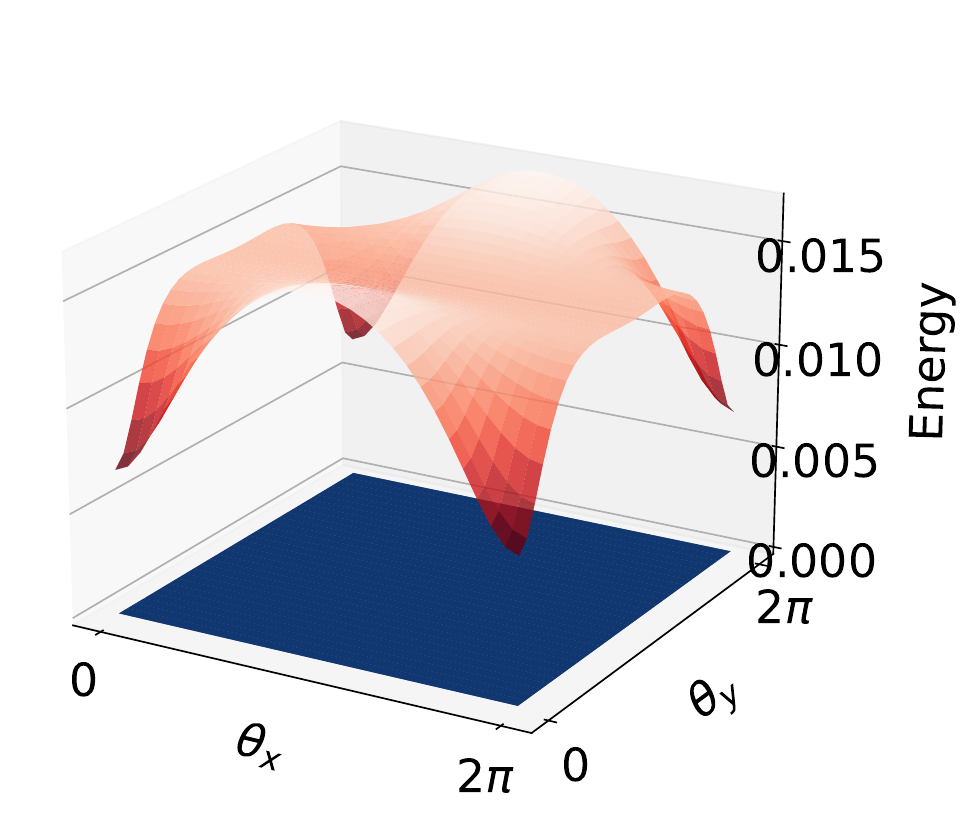}
        \centering
    \end{minipage}
    \hfill
    \begin{minipage}[b]{0.48\columnwidth}
        \includegraphics[width=\linewidth]{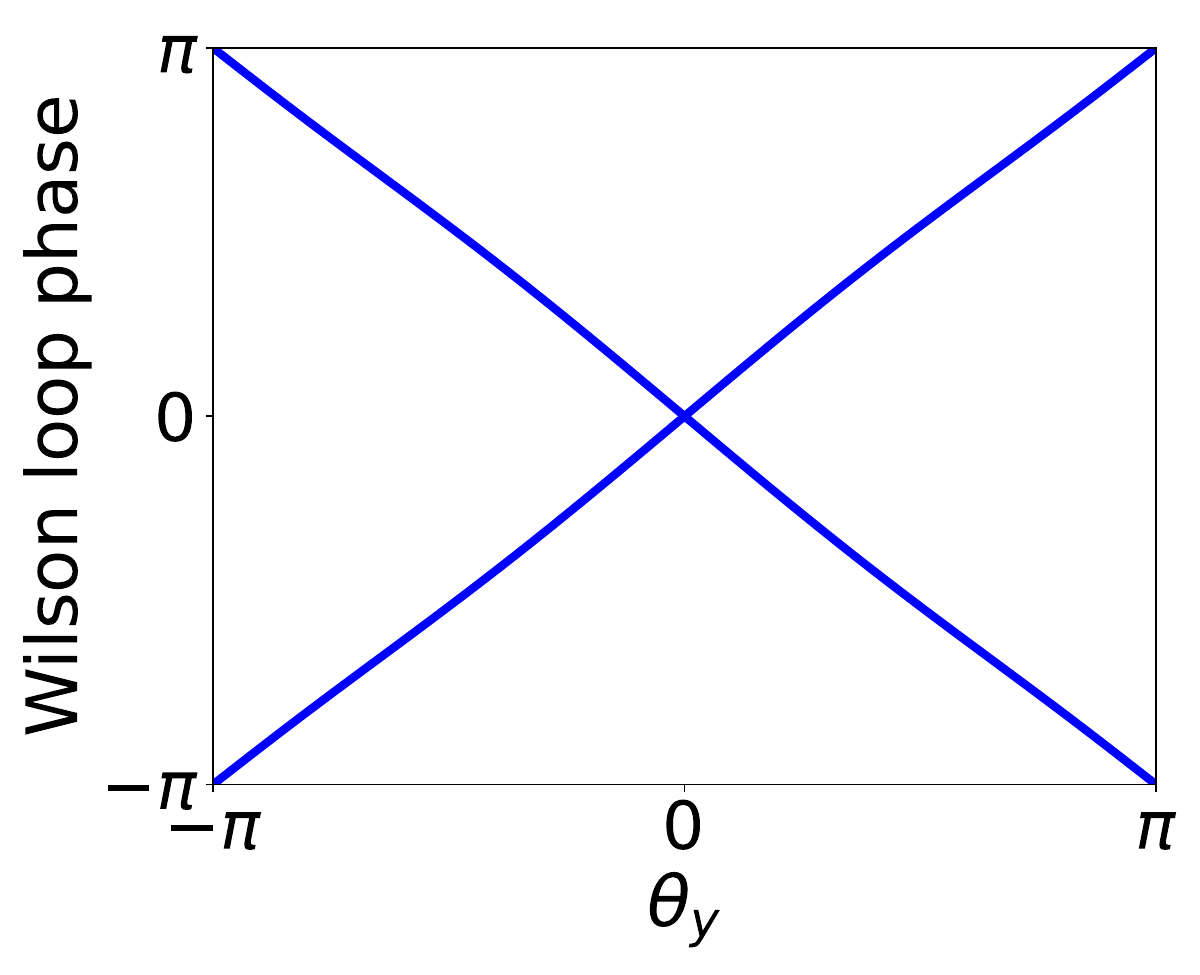}
        \centering
    \includegraphics[width=\linewidth]{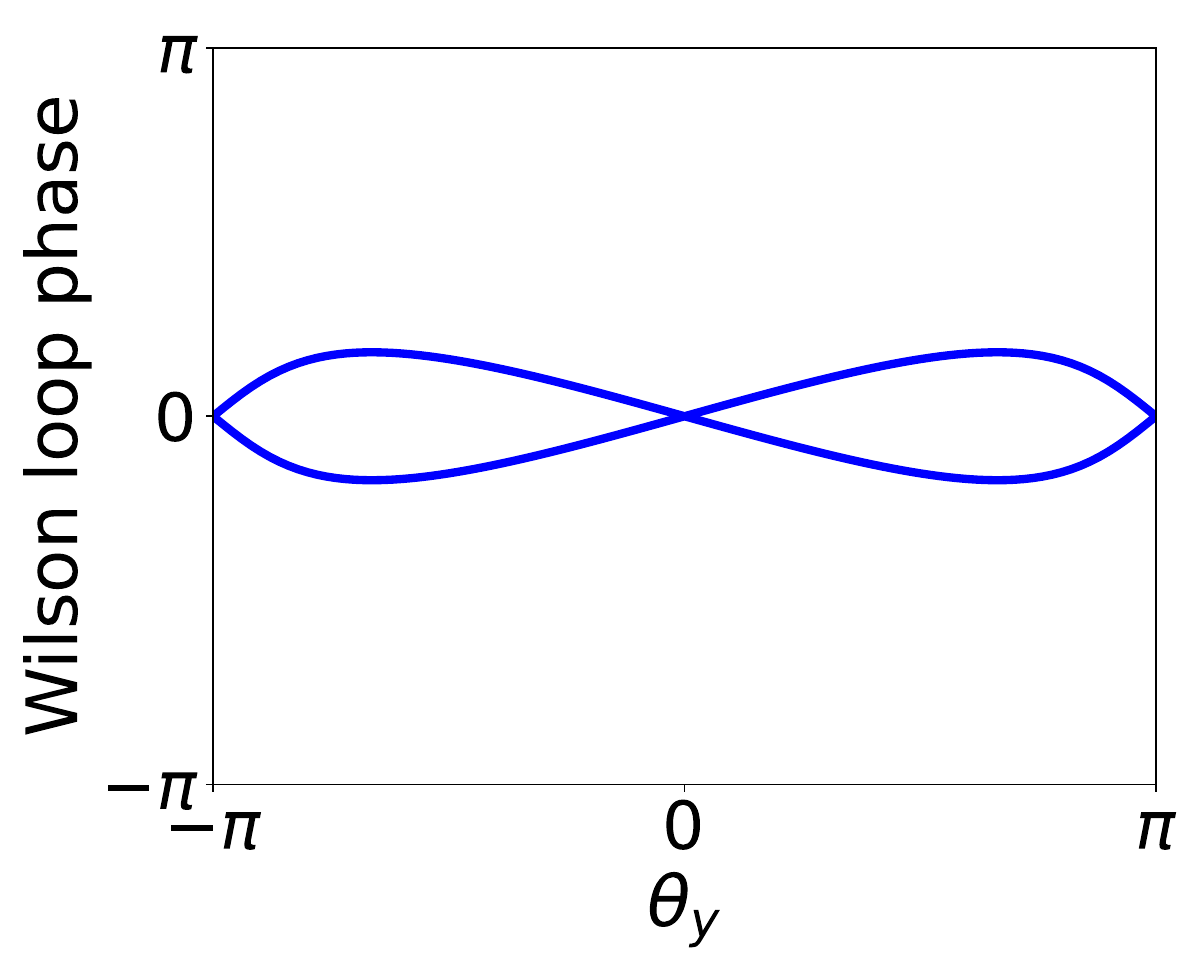}
        \centering
    \end{minipage}

    \caption{Left column: Low-energy spectrum of Eq.~\eqref{Eq:projected_Hamiltonian} with $t = 1, t' = 1/\sqrt{2}, \gamma = 0.1, U = 1$ (top) and for $t = 1, t' = 1/\sqrt{2}, \gamma = 0, U = 1$ with an onsite potential $\Delta = 2.5$ (bottom, see Eq.~\eqref{Eq:onsite}). For clarity, energies have been shifted so that the two (quasi)degenerate ground states (shown in blue) are centered around 0. In the plot with $\gamma = 0.1$, finite-size splittings between the ground states are smaller than the resolution of the figure. An energy gap protects the ground states from the lowest excitations (shown in red) for all $\vec{\theta}$. Darker shades indicate lower energy values. Right column: Wilson loop flows of the two ground states, computed using the same parameters as in the corresponding energy spectra shown in the left column.}
    \label{fig:AQHE}
\end{figure}

 We are, however, interested in detecting the ground state topology in absence of spin conservation ($\gamma\neq 0$). 
We solve the Hamiltonian in Eq.\eqref{Eq:projected_Hamiltonian} using exact diagonalization (ED) on a $C_4$-symmetric $4 \times 4$ cluster with 8 particles. 
The dependence on $\vec{\theta}$ enters the Hamiltonian, because now the projection operators in Eq.~\eqref{Eq:projected_Hamiltonian} depend on $\vec{\theta}$. We show the low-energy spectrum for $\gamma = 0.1$ in Fig.~\ref{fig:AQHE} (top left panel). As indicated by the Wilson loop flow in Fig.~\ref{fig:AQHE}~(top right panel) yielding $|e_2| = 1$, the addition of a small Rashba spin-orbit coupling does not alter the Hall conductivity. For comparison with a trivial charge density wave phase, we also performed a calculation with an onsite potential
\begin{equation}\label{Eq:onsite}
    \hat{H}_{\Delta} = \Delta \displaystyle\sum_{x,y} \displaystyle\sum_{\sigma} (-1)^{x + 1} c_{x, y, \sigma}^{\dagger} c_{x,y,\sigma},
\end{equation}
see Fig.~\ref{fig:AQHE} (bottom panels).

In the absence of other symmetries apart from $\hat{A} = \hat{C}_2 \hat{\mathcal{T}}$, the entire topology of the ground states is contained in $|e_2|$~\cite{Bou}. We can arrive at a much finer classification scheme by also considering crystalline symmetries, in particular rotation symmetries. If $\hat{g} \hat{H}(\vec{\theta}_{0}) \hat{g}^{-1} = \hat{H}(\vec{\theta}_{0})$ for a symmetry group $\lbrace \hat{g} \rbrace$, we can classify wave functions at $\vec{\theta}_{0}$ by irreducible representations (irreps) of the associated symmetry group. A wave function with different irreps then describes a topologically distinct phase as a gap closure is always required to arrive at different irreps. In particular, for the $4 \times 4$ cluster we can label wave functions at the high-symmetry points (HSP) $\vec{\theta} = (0,0), (\pi, \pi)$ by irreps of $C_4$ and for $\vec{\theta} = (0, \pi), (\pi, 0)$ by irreps of $C_2$. These irreps at HSPs  serve as symmetry indicators both for nontrivial topology in general and constraints on the Hall conductivity in particular, in analogy to single-particle band topology~\cite{Kru, Bra, Po2, Mat}. For our topological Hubbard model we find at $\vec{\theta} = (0,0)$ that the ground states transform under $E^{+} \oplus E^{-}$ and at the three other points under $2A$, indicating a Hall conductivity of $|\sigma_{xy}| = \frac{e^2}{h}$(1 mod 2)~\cite{Sup}. It is possible to extend this analysis by adding further crystalline symmetries such as mirror symmetry, to eventually give a complete set of possible symmetry indicators~\cite{Bir, Bou2}.  

\textit{Fractional Chern insulator.}---One of the hallmarks of fractional Chern insulators is topological ground state degeneracy associated to fractionally charged quasiparticles. Because of the enlarged degeneracy of the ground state, the Euler number is, in principle, mathematically no longer well defined~\footnote{Its parity $(-1)^{e_2}$ remains well-defined~\cite{Bou,Sup}, which would however fail to identify experimentally relevant FCI phases with $|\sigma_{xy}| = 2e^2/3h$~\cite{Zen, Wat}.}. We can circumvent this issue by using translation symmetry and grouping ground states into pairs with the same many-body momentum $K_x$. The emergence of states with different momenta does not originate from spontaneous translation symmetry breaking~\cite{Fun}, but instead is a result of center-of-mass degeneracy, which occurs on lattices with $N_e / N_x \neq 0$ mod 1, where $N_e$ is the particle number~\cite{Osh}. Since states with different $K_x$ cannot hybridize, we can partition the ground state Hilbert space into smaller subspaces.

\begin{figure}[t] 
    \centering
    \begin{minipage}[b]{0.48\columnwidth} 
        \includegraphics[width=\linewidth]{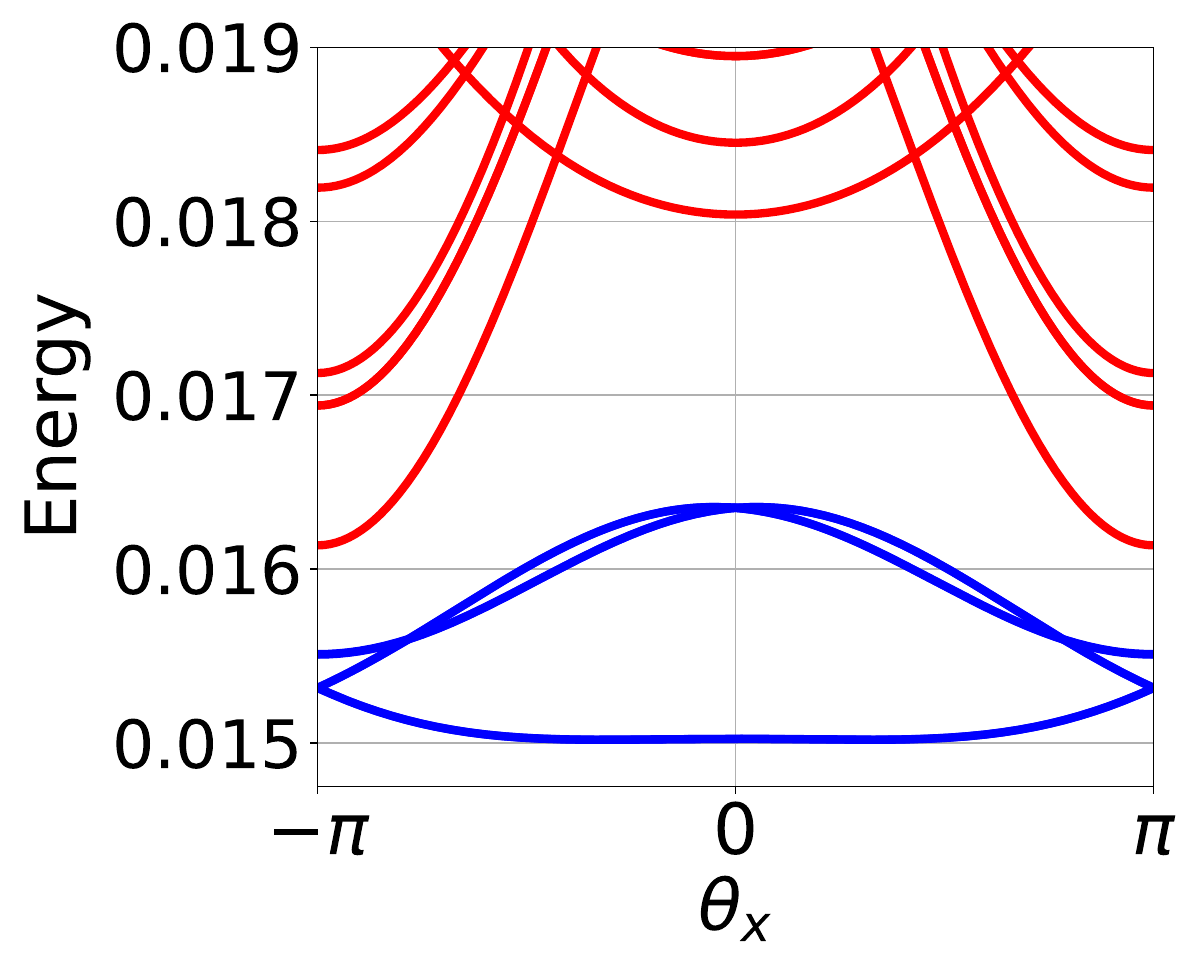}
        \centering
    \end{minipage}
    \hfill
    \begin{minipage}[b]{0.48\columnwidth}
        \includegraphics[width=\linewidth]{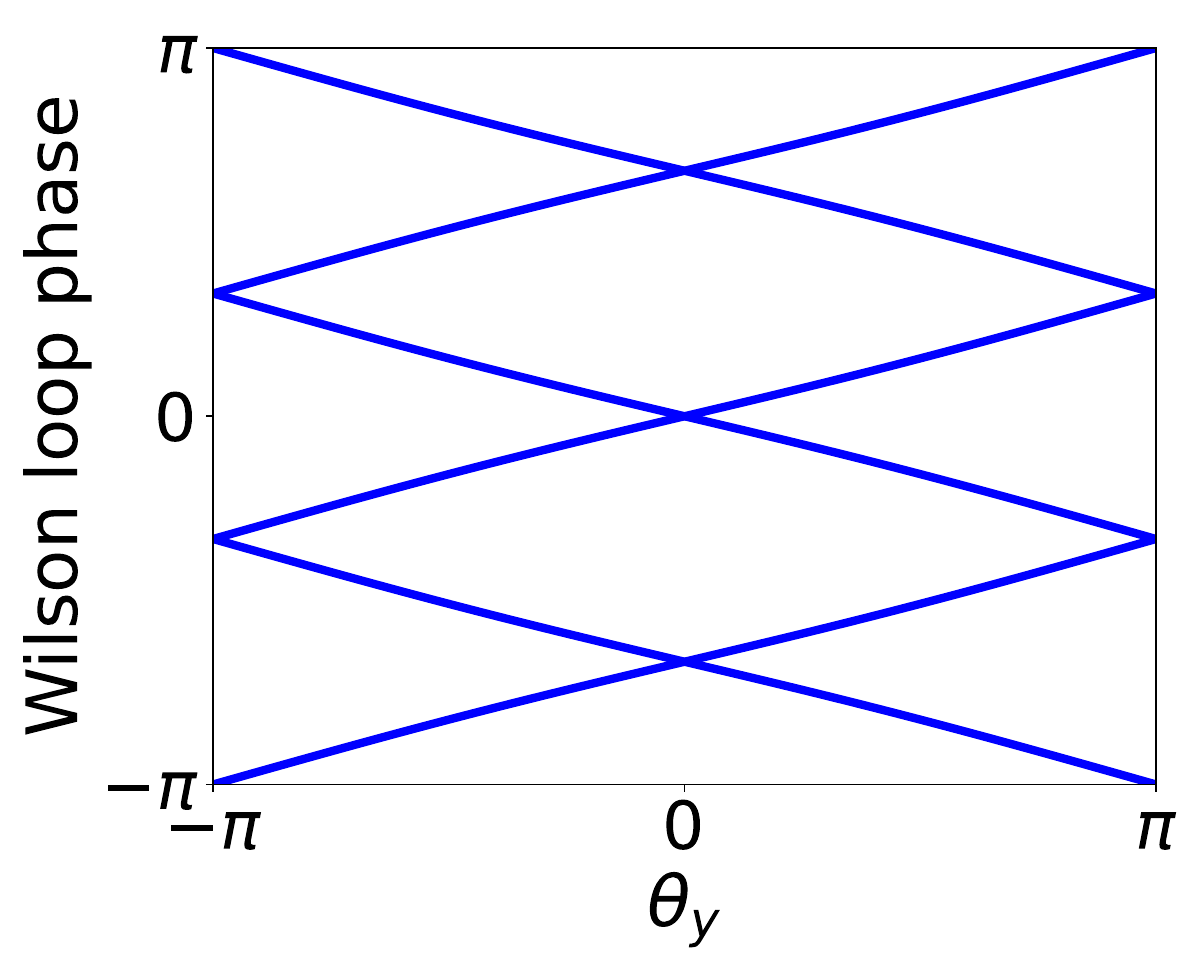}
        \centering
    \end{minipage}

    \caption{Left image: Low-energy spectrum of Eq.~\eqref{Eq:fractional_projected_hamiltonian} exhibiting spectral flow with $t = 1, t' = 1/\sqrt{2}, \gamma = 0.1, U = 1, V = 0.1, \theta_y = 0$. The finite-size splitting of the (quasi)degenerate ground states (shown in blue) with the same momentum is smaller than the linewidth. The excitation spectrum (shown in red) is gapped from the ground state spectrum for all $\vec{\theta}$. Right image: Wilson loop flow for the ground states in the fractional Chern insulating phase.}
    \label{fig:FQHE}
\end{figure}

On a $6 \times 4$ cluster with 4 particles, if we include a nearest-neighbor Coulomb interaction term
\begin{equation}
    \hat{H}_{V} = V \displaystyle\sum_{x,y} ( \hat{n}_{x,y} \hat{n}_{x+1,y} + \hat{n}_{x,y} \hat{n}_{x,y+1})
\end{equation}
so that the Hamiltonian becomes
\begin{equation}\label{Eq:fractional_projected_hamiltonian}
    \hat{H}(\vec{\theta}) = \hat{P}(\vec{\theta}) (\hat{H}_{U} + \hat{H}_{V}) \hat{P}(\vec{\theta}).
\end{equation}
We can stabilize a spin-polarized, fractional Chern insulating solution. Its six ground states come in pairs of momenta $(K_x, K_y) = (0, \pi), (2\pi/3, \pi), (4\pi/3,\pi)$ (note that $K_x$ corresponds to two lattice site translations due to the enlarged unit cell). This leads to a well-defined Euler number. We plot the low-energy and Wilson loop flows in Fig.~\ref{fig:FQHE}. We set $\gamma = 0.02$ to demonstrate that the winding number is well defined in the absence of $S^{z}$ conservation. The Wilson loop operator winds once every three cycles, indicating a fractional Hall conductivity of $|\sigma_{xy}| = e^2/3h$. Analogously to the Chern insulating case, the nontrivial topology can also be read off by symmetry indicators~\footnote{At $\vec{\theta} = (\pi, 0)$ we find the ground state wave functions to transform like $2A \oplus 4 B$ and at the remaining $C_2$ invariant points to transform like $4A \oplus 2B$, indicating $\sigma_{xy} = e^2/3h$ mod $2e^2/3h$.}.

\textit{Conclusion and Outlook.}---In this paper we have introduced many-body Euler numbers as a new topological invariant characterizing the spontaneous breaking of time-reversal invariance in Chern and fractional Chern insulators. Considering crystalline symmetries, we sketched how this approach can be extended to distinguish topologically inequivalent phases, in analogy to classifying schemes in single-particle band topology. Using exact diagonalization, we calculated the Hall conductivity of Chern and fractional Chern insulating phases, as well as a trivial charge-density wave phase of a topological Hubbard model exemplarily.

Regarding applications for real materials, our method may be employed to model twisted bilayer structures or transition metal dichalcogenides, where the approximate spin and valley symmetries -- often invoked to disentangle time-reversal partners -- may be lifted by strain~\cite{RefAi, RefZha} or Rashba spin-orbit coupling~\cite{RefZol, RefYao, RefHu, RefFen}, as well as altermagnetic or noncollinear magnetic insulators~\cite{RefJia, RefLiu}. Furthermore, at the current pace of progress~\cite{RefAid, RefJot, RefLon, RefZho} in cold-atom systems, they are likely to emerge as relevant experimental platforms in the near future, particularly in light of recent realizations of bosonic analogs of the anomalous quantum Hall effect with spontaneously broken time-reversal invariance~\cite{RefWan2}.

\begin{acknowledgments}

This work was supported by the Deutsche Forschungsgemeinschaft (DFG, Project No.~449872909) and the Swiss National Science Foundation (Project No. 200021E\_198011) through QUAST-FOR5249  (projects TP3 and TP4).

\end{acknowledgments}

\textit{Data availability.}---Code and data points for the figures are publicly available~\cite{data}.

\bibliography{main_bib}

\clearpage


\section*{Supplementary material}

# Supplementary Material for: Many-body Euler Topology

Axel Fünfhaus,[1,][*] Titus Neupert,[2] Thilo Kopp,[3] and Roser Valentí[1]

[1]*Institute of Theoretical Physics, Goethe University Frankfurt,*
*Max-von-Laue-Straße 1, 60438 Frankfurt am Main, Germany*
[2]*Department of Physics, University of Zurich, Winterthurerstrasse 190, 8057 Zurich, Switzerland*
[3]*Center for Electronic Correlations and Magnetism, Experimental Physics VI,*
*Institute of Physics, University of Augsburg, 86135 Augsburg, Germany*

(Dated: January 20, 2026)

## Appendix A: Single-particle Hamiltonians coupling to fluxes

### 1. Eigenspectra

It is possible to pierce the two noncontractable loops of a finite cluster subject to periodic boundary conditions with fluxes $\Phi_x$ and $\Phi_y$. A particle hopping around such a loop must obtain a phase $\theta_{x/y} = -2\pi \Phi_{x/y}/\Phi_0$, where $\Phi_0 = hc/e$ is the magnetic flux quantum. Crucially, only phases originating from particles hopping around closed loops and the magnitudes of the hoppings are gauge invariant [1]. One can therefore pick the "twisted boundary gauge", where particles gain a phase factor $e^{-i\theta_{x/y}}$ if they hop from the unit cell at $x/y = N_{x/y}$ to $x/y = 1$, where $N_x$ is the number of unit cells in $x$ direction and $N_y$ is the number of unit cells in y direction.

Now, consider a general class of translation-invariant hopping Hamiltonians that we parameterize with a twist angle, so that $\hat{H} = \hat{H}(\theta_x, \theta_y) = \hat{H}(\theta_x + 2\pi, \theta_y) = \hat{H}(\theta_x, \theta_y + 2\pi)$. To diagonalize the single-particle Hamiltonian with twisted boundary conditions, it is expedient to work with the Fourier transform of creation operators in real space. To have no $\boldsymbol{\theta}$ dependence in the translation operators, we perform a gauge transformation on the Hamiltonian to make it genuinely translation invariant [2]. In general, a gauge transformation has the form[1]

$$\hat{\lambda} = e^{-i \sum_m \lambda_m c_m^\dagger c_m}$$
$$\hat{\lambda} c_m \hat{\lambda}^\dagger = c_m e^{i\lambda_m}$$
$$\hat{\lambda} c_m^\dagger \hat{\lambda}^\dagger = e^{-i\lambda_m} c_m^\dagger. \quad (A1)$$

To transform from the twisted boundary gauge to the translation invariant "periodic gauge", we identify

$$\hat{\lambda}(\boldsymbol{\theta}) = e^{-i \sum_{(m,n),\alpha} m \frac{\theta_x}{N_x} c_{(m,n),\alpha}^\dagger c_{(m,n),\alpha}}$$
$$\cdot e^{-i \sum_{(m,n),\alpha} n \frac{\theta_y}{N_y} c_{(m,n),\alpha}^\dagger c_{(m,n),\alpha}}$$
$$= e^{-i \sum_{(m,n),\alpha} \lambda_{(m,n)}(\boldsymbol{\theta}) c_{(m,n),\alpha}^\dagger c_{(m,n),\alpha}}, \quad (A2)$$

---

[*] fuenfhaus@itp.uni-frankfurt.de
[1] In the following, operators acting on the physical Hilbert space will be denoted with a hat, in contrast to base-dependent matrix representations of operators. For convenience, we omit the hat for creation and annihilation operators.

where $(m, n)$ refers to the unit cell, $\alpha$ is a sublattice basis (including spin) and $N_x, N_y$ are the number of unit cells in $x$ and $y$ direction. A generic hopping Hamiltonian in twisted boundary gauge has the form

$$\hat{H}(\boldsymbol{\theta}) = \sum_{\substack{m,n \\ m',n'}}^{N_x, N_y} \sum_{\alpha,\beta}^{N_{sub}} c_{(m,n),\alpha}^\dagger h_{(m,n),\alpha;(m',n'),\beta}(\boldsymbol{\theta}) c_{(m',n'),\beta}, \quad (A3)$$

with the number of sublattice degrees of freedom $N_{sub}$. We define

$$\hat{\tilde{H}}(\boldsymbol{\theta}) = \hat{\lambda}(\boldsymbol{\theta}) \hat{H}(\boldsymbol{\theta}) \hat{\lambda}(\boldsymbol{\theta})^\dagger \quad (A4)$$

and

$$\tilde{h}_{(m,n),\alpha;(m',n'),\beta}(\boldsymbol{\theta})$$
$$= e^{-i\lambda_{(m,n)}(\boldsymbol{\theta})} e^{i\lambda_{(m',n')}(\boldsymbol{\theta})} h_{(m,n),\alpha;(m',n'),\beta}(\boldsymbol{\theta}). \quad (A5)$$

By construction, $\tilde{h}$ is translation invariant, so

$$\tilde{h}_{(m,n),\alpha;(m',n'),\beta}(\boldsymbol{\theta})$$
$$= \tilde{h}_{((m-m'+1) \bmod N_x, (n-n'+1) \bmod N_y),\alpha;(1,1),\beta}(\boldsymbol{\theta})$$
$$= \tilde{h}_{((m-m') \bmod N_x, (n-n') \bmod N_y),\alpha,\beta}(\boldsymbol{\theta}) \quad (A6)$$

(where we will omit in the following the mod $N_{x/y}$ for simplicity) and the gauge-transformed Hamiltonian becomes

$$\hat{\tilde{H}}(\boldsymbol{\theta}) = \sum_{\substack{m,n \\ m',n'}}^{N_x, N_y} \sum_{\alpha,\beta}^{N_{sub}} c_{(m,n),\alpha}^\dagger \tilde{h}_{(m-m',n-n'),\alpha,\beta}(\boldsymbol{\theta}) c_{(m',n'),\beta}. \quad (A7)$$

To diagonalize the Hamiltonian, we first Fourier-transform the creation operators

$$c_{(m,n),\alpha} = \frac{1}{\sqrt{N}} \sum_{\boldsymbol{k}} e^{i(m k_x + n k_y)} c_{(k_x,k_y),\alpha}. \quad (A8)$$

One then readily finds

$$\hat{\tilde{H}}(\boldsymbol{\theta}) = \sum_{\boldsymbol{k}} \sum_{\alpha,\beta} \tilde{h}_{\alpha,\beta}(\boldsymbol{k}, \boldsymbol{\theta}) c_{\boldsymbol{k},\alpha}^\dagger c_{\boldsymbol{k},\beta}, \quad (A9)$$

with

$$\tilde{h}_{\alpha,\beta}(\boldsymbol{k},\boldsymbol{\theta}) = \sum_{\bar{m},\bar{n}=0}^{N_x-1,N_y-1} \tilde{h}_{(\bar{m},\bar{n}),\alpha,\beta}(\boldsymbol{\theta})e^{-ik_x\bar{m}}e^{-ik_y\bar{n}}$$
$$= \sum_{\bar{m},\bar{n}=0}^{N_x-1,N_y-1} e^{-i\bar{m}(k_x+\theta_x/N_x)-i\bar{n}(k_y+\theta_y/N_y)}$$
$$\cdot h_{(\bar{m}+1,\bar{n}+1),\alpha;(1,1),\beta}(\boldsymbol{\theta}). \quad \text{(A10)}$$

Since

$$e^{i\theta_x} = e^{iN_x(k_x+\theta_x/N_x)}, \quad \text{(A11)}$$

as $k_x = 2\pi j/N_x$ with $j$ a natural number and similarly for $k_y$, we finally identify

$$\hat{\tilde{H}}(\boldsymbol{\theta}) = \sum_{\boldsymbol{k}}\sum_{\alpha,\beta} \tilde{h}_{\alpha,\beta}(k_x+\theta_x/N_x,k_y+\theta_y/N_y)c^\dagger_{\boldsymbol{k},\alpha}c_{\boldsymbol{k},\beta}$$
$$= \sum_{\boldsymbol{k}}\sum_{\alpha,\beta} \tilde{h}_{\alpha,\beta}(\boldsymbol{k_\theta})c^\dagger_{\boldsymbol{k},\alpha}c_{\boldsymbol{k},\beta}, \quad \text{(A12)}$$

where

$$\tilde{h}_{\alpha,\beta}(k_x+\theta_x/N_x,k_y+\theta_y/N_y) = \tilde{h}_{\alpha,\beta}(\boldsymbol{k_\theta}). \quad \text{(A13)}$$

### 2. Chern numbers

We see that the single-particle modes get shifted by $\boldsymbol{\theta}$. We now want to recover the expression of the Hall-conductivity via an integral of the Berry-curvature in the Brillouin zone. The eigenstates for a band insulator are given by

$$|\Psi(\boldsymbol{\theta})\rangle = \prod_{\boldsymbol{k}}\prod_{1\le i\le N_{\text{occ}}} \tilde{c}^\dagger_{\boldsymbol{k_\theta},i}|0\rangle, \quad \text{(A14)}$$

where $N_{\text{occ}}$ is the number of occupied bands and we stipulate for the eigenenergies $\{\epsilon_i(\boldsymbol{k_\theta})\}$ that they are ordered according to magnitude. The single-particle eigenmodes are defined by[2]

$$\tilde{c}^\dagger_{\boldsymbol{k_\theta},i} = \left(U^\dagger_{\boldsymbol{k_\theta}}\right)_{\alpha i} c^\dagger_{\boldsymbol{k},\alpha}. \quad \text{(A15)}$$

---

[2] Let a single-particle Hamiltonian be defined as

$$\hat{h} = \sum_{i,j} h_{ij}c^\dagger_i c_j, \quad U\hat{h}U^\dagger = \hat{\epsilon}$$

and its eigenstates as

$$|\epsilon_\alpha\rangle = \sum_j v^{\epsilon_\alpha}_j c^\dagger_j|0\rangle.$$

One then readily identifies

$$\hat{h}\boldsymbol{v}^{\epsilon_\alpha} = \epsilon_\alpha \boldsymbol{v}^{\epsilon_\alpha}, \quad U^\dagger = (\boldsymbol{v}^1,\boldsymbol{v}^2,\dots).$$

For the Berry curvature we fist evaluate the Berry connection:

$$\boldsymbol{A}(\boldsymbol{\theta}) = i\langle\Psi(\boldsymbol{\theta})|\boldsymbol{\nabla_\theta}|\Psi(\boldsymbol{\theta})\rangle. \quad \text{(A16)}$$

Using the product rule we find

$$\boldsymbol{A}(\boldsymbol{\theta}) = i\sum_{\boldsymbol{k}}\sum_{1\le i\le N_{\text{occ}}} \langle 0|\tilde{c}_{\boldsymbol{k_\theta},i}\boldsymbol{\nabla_\theta}\tilde{c}^\dagger_{\boldsymbol{k_\theta},i}|0\rangle$$
$$= i\sum_{\boldsymbol{k}}\sum_{1\le i\le N_{\text{occ}}}\sum_{\alpha,\beta} \langle 0|c_{\boldsymbol{k},\alpha}(U_{\boldsymbol{k_\theta}})_{i\alpha}\boldsymbol{\nabla_\theta}\left(U^\dagger_{\boldsymbol{k_\theta}}\right)_{\beta i}c^\dagger_{\boldsymbol{k},\beta}|0\rangle$$
$$= i\sum_{\boldsymbol{k}}\sum_{1\le i\le N_{\text{occ}}}\sum_{\alpha} (U_{\boldsymbol{k_\theta}})_{i\alpha}\boldsymbol{\nabla_\theta}\left(U^\dagger_{\boldsymbol{k_\theta}}\right)_{\alpha i}. \quad \text{(A17)}$$

For the Chern number [3] we need to integrate the Berry curvature over the square $D^2 = [0,2\pi]\times[0,2\pi]$:

$$C = \frac{1}{2\pi}\int_{D^2} d\theta_x\,d\theta_y$$
$$\boldsymbol{\nabla_\theta}\times\left\{i\sum_{\boldsymbol{k}}\sum_{1\le i\le N_{\text{occ}}}\sum_\alpha (U_{\boldsymbol{k_\theta}})_{i\alpha}\boldsymbol{\nabla_\theta}\left(U^\dagger_{\boldsymbol{k_\theta}}\right)_{\alpha i}\right\}. \quad \text{(A18)}$$

Note that, other than in twisted boundary gauge, $D^2$ does not form a torus. The Berry-curvature is not gauge-invariant under gauge transformations in real space (like switching from twisted boundary gauge to the periodic gauge), but the topological invariants do not change, justifying Eq. (A18) [4, 5].

We want to replace the integral over $\boldsymbol{\theta}$ with an integral over $\boldsymbol{k}$. We first substitute the integral over $\boldsymbol{\theta}$ with an integral over $\boldsymbol{k_\theta}$:

$$C = \frac{i}{2\pi}\sum_{\boldsymbol{k}}\sum_{1\le i\le N_{\text{occ}}}\int_{\boldsymbol{k}+D^2_{N_x,N_y}} d^2\boldsymbol{k_\theta}\boldsymbol{\nabla_{k_\theta}}$$
$$\times\left\{\sum_\alpha (U_{\boldsymbol{k_\theta}})_{i\alpha}\boldsymbol{\nabla_{k_\theta}}\left(U^\dagger_{\boldsymbol{k_\theta}}\right)_{\alpha i}\right\}, \quad \text{(A19)}$$

where $D^2_{N_x,N_y}$ is $[0,2\pi/N_x]\times[0,2\pi/N_y]$. By evaluating the sum over $\boldsymbol{k}$ we find

$$C = \frac{i}{2\pi}\int_{T^2} d^2\boldsymbol{k}\boldsymbol{\nabla_k}\times\text{Tr}(\boldsymbol{A}(\boldsymbol{k})), \quad \text{(A20)}$$

where $\boldsymbol{A}$ is the non-Abelian Berry connection defined by

$$(\boldsymbol{A})_{ij} = \sum_\alpha (\boldsymbol{v}_{\boldsymbol{k},i})^*_\alpha \boldsymbol{\nabla_k}(\boldsymbol{v}_{\boldsymbol{k},j})_\alpha, \quad \text{(A21)}$$

and $\boldsymbol{v}_{\boldsymbol{k},i}$ are the single-particle eigenstates of $\tilde{h}(\boldsymbol{k})$. This is exactly the same expression as in the single-particle case.

### 3. Euler numbers

We finally take a look at the Euler number for time-reversal invariance-breaking ground states that can be

written as product states of Bloch waves. For a band insulator the Euler number is defined as [6, 7]

$$e_2 = \frac{1}{2\pi} \int_{BZ} d^2\boldsymbol{k} \mathcal{F}_{+-}(\boldsymbol{k}), \quad (A22)$$

with

$$\mathcal{F}_{+-}(\boldsymbol{k}) = \boldsymbol{\nabla}_{\boldsymbol{k}} \times \boldsymbol{\mathcal{A}}_{+-}(\boldsymbol{k}), \qquad \boldsymbol{\mathcal{A}}_{+-}(\boldsymbol{k}) = \left(\boldsymbol{v}_{\boldsymbol{k}}^{+}\right)^T \boldsymbol{\nabla}_{\boldsymbol{k}} \boldsymbol{v}_{\boldsymbol{k}}^{-}, \quad (A23)$$

where $\boldsymbol{v}_{\boldsymbol{k}}^{\pm}$ are time-reversal invariant or real states (i.e. their vector elements are real numbers). In the case of the anomalous quantum Hall effect we have two many-body states, one with $C=1$, one with $C=-1$ (the generalization to different Chern numbers is trivial). We then have two states that are partners[3] under $\hat{A} = C_2 \mathcal{T}$:

$$|\Psi_{C=1}(\boldsymbol{\theta})\rangle = \prod_{\boldsymbol{k}} \tilde{c}_{\boldsymbol{k}_{\boldsymbol{\theta}},C=1}^{\dagger}|0\rangle$$

$$|\Psi_{C=-1}(\boldsymbol{\theta})\rangle = \prod_{\boldsymbol{k}} \tilde{c}_{\boldsymbol{k}_{\boldsymbol{\theta}},C=-1}^{\dagger}|0\rangle$$

$$|\Psi_{+}(\boldsymbol{\theta})\rangle = \frac{1}{\sqrt{2}} \left(|\Psi_{C=1}(\boldsymbol{\theta})\rangle + |\Psi_{C=-1}(\boldsymbol{\theta})\rangle\right)$$

$$|\Psi_{-}(\boldsymbol{\theta})\rangle = \frac{1}{\sqrt{2}} \left(i|\Psi_{C=1}(\boldsymbol{\theta})\rangle - i|\Psi_{C=-1}(\boldsymbol{\theta})\rangle\right). \quad (A24)$$

It is evident that

$$\langle \Psi_{+}(\boldsymbol{\theta}) | \boldsymbol{\nabla}_{\boldsymbol{\theta}} | \Psi_{-}(\boldsymbol{\theta}) \rangle = 0, \quad (A25)$$

because, due to the product rule, $\boldsymbol{\nabla}_{\boldsymbol{\theta}}$ acts only on a single creation operator with some $\boldsymbol{k}$ in each summand; due to the difference in the remaining creation operators at other $\boldsymbol{k}'$ in the two product states, Eq. A25 has to vanish identically. We are then left with

$$e_2 = \frac{1}{4\pi} \int_{T^2} d^2\boldsymbol{\theta} \boldsymbol{\nabla}_{\boldsymbol{\theta}} \times \{i\langle \Psi_{C=1}(\boldsymbol{\theta})|\boldsymbol{\nabla}_{\boldsymbol{\theta}}|\Psi_{C=1}(\boldsymbol{\theta})\rangle$$
$$- i\langle \Psi_{C=-1}(\boldsymbol{\theta})|\boldsymbol{\nabla}_{\boldsymbol{\theta}}|\Psi_{C=-1}(\boldsymbol{\theta})\rangle\}$$
$$= 1. \quad (A26)$$

We see that the anomalous Hall transport is correctly captured by the Euler number.

### Appendix B: Wilson loops

#### 1. Basics of Wilson loop operators

Wilson loop operators [8] can be thought of as geometric components of time evolution operators that capture geometric aspects of the problem at hand. In the following we consider a system with $n$ ground states that are protected by an energy gap from excited states. For that we make in the context of the adiabatic theorem [9] the approximation that we can write the time-dependent Hamiltonian as[4]

$$\hat{H}(\boldsymbol{\mathcal{R}}(t)) = \hat{V}(\boldsymbol{\mathcal{R}}(t))\hat{H}_0\hat{V}(\boldsymbol{\mathcal{R}}(t))^{\dagger} \quad (B1)$$

and we assume that the ground states are exactly degenerate and have eigenenergy 0. Ground states are then states

$$\hat{V}(\boldsymbol{\mathcal{R}}(t))|\Psi_m\rangle, \quad (B2)$$

where $\{|\Psi_m\rangle\}$ are the ground states at $t_0$. To evaluate this expression for a closed path in parameter space $\boldsymbol{\mathcal{R}}$ we first partition the time evolution operator:

$$\hat{U}(t, t_0) = \lim_{N \to \infty} \hat{U}(t, t_1)\hat{U}(t_1, t_2)\ldots\hat{U}(t_N, t_0), \quad (B3)$$

where $t_{i+1} - t_i \ll 1$. We identify

$$\hat{U}(t_i, t_{i+1}) = \hat{T} \exp\left(-\frac{i}{\hbar} \int_{t_i}^{t_{i+1}} dt' \hat{H}(t')\right)$$
$$\approx \hat{V}(\boldsymbol{\mathcal{R}}(t_{i+1})) \exp\left(-\frac{i}{\hbar}\Delta t \hat{H}_0\right) \hat{V}(\boldsymbol{\mathcal{R}}(t_{i+1}))^{\dagger}$$
$$\approx \hat{V}(\boldsymbol{\mathcal{R}}(t_{i+1})) \left(1 - \frac{i}{\hbar}\Delta t \hat{H}_0\right) \hat{V}(\boldsymbol{\mathcal{R}}(t_{i+1}))^{\dagger}. \quad (B4)$$

If we apply the time evolution operator only onto the ground states then $\hat{V}$ will approximately only transform the ground states into themselves and all the terms containing $\hat{H}_0$ vanish. More precisely, let $\hat{P}_{GS}$ be the projector onto ground states and $\hat{P}_{Exc}$ the projector onto excited states of $\hat{H}_0$:

$$\left(1 - \frac{i}{\hbar}\Delta t \hat{H}_0\right) = \hat{P}_{GS} + \hat{P}_{Exc}\left(1 - \frac{i}{\hbar}\Delta t \hat{H}_0\right)\hat{P}_{Exc}. \quad (B5)$$

Since $\hat{P}_{GS}\hat{V}(t_i)^{\dagger}\hat{V}(t_{i+1})\hat{P}_{Exc} \approx 0$, in Eq. (B3) we are then left with terms

$$\hat{P}_{GS}\hat{V}(\boldsymbol{\mathcal{R}}(t_i))^{\dagger}\hat{V}(\boldsymbol{\mathcal{R}}(t_{i+1}))\hat{P}_{GS}. \quad (B6)$$

To evaluate them, it will be useful to approximate the Berry connection (below we use the Einstein sum convention):

$$\boldsymbol{A}_{mn}(\boldsymbol{\mathcal{R}}(t)) = i\langle \Psi_m|\hat{V}^{\dagger}(\boldsymbol{\mathcal{R}}(t))\boldsymbol{\nabla}_{\boldsymbol{\mathcal{R}}}\hat{V}(\boldsymbol{\mathcal{R}}(t))|\Psi_n\rangle$$
$$\approx i\langle \Psi_m|\hat{V}^{\dagger}(\boldsymbol{\mathcal{R}}(t))$$
$$\cdot \boldsymbol{v}_i\left(\frac{\hat{V}(\boldsymbol{\mathcal{R}}(t) + \boldsymbol{v}_i\Delta\mathcal{R}_i) - \hat{V}(\boldsymbol{\mathcal{R}}(t))}{\Delta\mathcal{R}_i}\right)|\Psi_n\rangle$$
$$= -i\frac{\boldsymbol{v}_i}{\Delta\mathcal{R}_i}\delta_{mn}$$
$$+ i\frac{\boldsymbol{v}_i}{\Delta\mathcal{R}_i}\langle \Psi_m(\boldsymbol{\mathcal{R}}(t))|\Psi_n(\boldsymbol{\mathcal{R}}(t) + \boldsymbol{v}_i\Delta\mathcal{R}_i)\rangle$$
$$(B7)$$

---

[3] Here, we assume that $\hat{A}$ does not have a $\boldsymbol{\theta}$ dependence, which is the case for the cluster considered in the main text.

[4] It is clear that this an approximation. For example, deviation of the eigenenergy of any state over time cannot be captured with this simplifying ansatz.

Let $\Delta\mathcal{R} = \Delta\mathcal{R}_\alpha \boldsymbol{v}_\alpha$ and $\mathcal{R}(t + \Delta t) \approx \mathcal{R}(t) + \Delta\mathcal{R}$. Then Eq. (B7) becomes

$$\boldsymbol{A}_{mn}(\mathcal{R}(t))\Delta\mathcal{R} \approx -i\delta_{mn} + i\langle\Psi_m(\mathcal{R}(t))|\Psi_n(\mathcal{R}(t+\Delta t))\rangle, \tag{B8}$$

so

$$\boldsymbol{A}(\mathcal{R}(t_i))\Delta\mathcal{R} = \langle -i\hat{1} + i\hat{V}(\mathcal{R}(t_i))^\dagger \hat{V}(\mathcal{R}(t_{i+1}))\rangle_{GS}, \tag{B9}$$

where $\langle...\rangle_{GS}$ is the expectation value over the ground states at $t_0$. Finally, to evaluate the time evolution operator we make the convention that $\hat{V}(\mathcal{R}(t_0)) = \hat{1}$, so we do not need to "reshuffle" the ground states. Then, if we have a closed path in parameter space so that the Hamiltonian maps back to its original self (this is the case in twisted boundary gauge, but not in periodic gauge, see below) we also stipulate that $\hat{V}(\mathcal{R}(t)) = \hat{1}$. Then

$$\begin{aligned}
&\langle \hat{U}(t,t_0)\rangle_{GS} \\
&= \lim_{N\to\infty}(1 - i\Delta\mathcal{R}(t_1)\boldsymbol{A}(\mathcal{R}(t_1)))\ldots \\
&\quad\ldots(1 - i\Delta\mathcal{R}(t_{N-1})\boldsymbol{A}(\mathcal{R}(t_{N-1}))) \\
&= \lim_{N\to\infty} e^{-i\Delta\mathcal{R}(t_1)\boldsymbol{A}(\mathcal{R}(t_1))}\ldots \\
&\quad\ldots e^{-i\Delta\mathcal{R}(t_{N-1})\boldsymbol{A}(\mathcal{R}(t_{N-1}))} \\
&= \hat{P}_{\text{Path ordering}} \exp\left(-i\oint_\mathcal{C} \boldsymbol{A}d\mathcal{R}\right),
\end{aligned} \tag{B10}$$

where $\mathcal{C}$ is the closed path and $\hat{P}_{\text{Path ordering}}$ orders the exponential factors. For clarity we note that this path ordering "operator" has nothing to do with the path integral formalism. The expression in Eq. (B10) is called Wilson loop operator and can be interpreted as a time evolution operator acting on ground states for adiabatically weak perturbations of the Hamiltonian.

How can we evaluate the Wilson loop operator numerically? Acting on the ground state we readily find using Eq. (B9)

$$W[\mathcal{C}] = \lim_{N\to\infty} F_0 \ldots F_N, \tag{B11}$$

with

$$\begin{aligned}
(F_j)_{mn} &= \langle\Psi_m(\mathcal{R}(t_j)|\Psi_n(\mathcal{R}(t_{j+1}))\rangle \\
&=: \langle\Psi_m(\mathcal{R}_j)|\Psi_n(\mathcal{R}_{j+1})\rangle.
\end{aligned} \tag{B12}$$

This expression is gauge invariant (see next section) we define $t_{N+1} = t$ and we have the convention that since $\mathcal{R}(t_0) = \mathcal{R}(t)$ we have $|\Psi_m(\mathcal{R}(t_0))\rangle = |\Psi_m(\mathcal{R}(t))\rangle$.

### 2. Gauge (in)dependence of Wilson loop operators

Regarding gauge (in)dependence there are two different questions to be answered: First, do we need to calculate the eigenstates in a gauge, where the Hamiltonian matrix is real and second, is there a difference between computing the Wilson loop operator in gauges other than the twisted boundary gauge?

First, we assume that the Hamiltonian matrix is real. Let $S$ be a unitary transformation mapping the ground states onto other ground states. We will show that the eigenspectrum of the Wilson loop operator is invariant. Let

$$|\Psi_\alpha(\mathcal{R}_{j+1})\rangle \to |\Psi_\beta(\mathcal{R}_{j+1})\rangle S^\dagger_{\beta\alpha}. \tag{B13}$$

The Wilson loop operator is invariant under this transformation for $j \neq -1, N$, because

$$\begin{aligned}
(F_j F_{j+1})_{mn} \\
= \sum_\alpha \langle\Psi_m(\mathcal{R}_j)|\Psi_\alpha(\mathcal{R}_{j+1})\rangle\langle\Psi_\alpha(\mathcal{R}_{j+1})|\Psi_n(\mathcal{R}_{j+2})\rangle
\end{aligned} \tag{B14}$$

and

$$\begin{aligned}
&\sum_\alpha |\Psi_\alpha(\mathcal{R}_{j+1})\rangle\langle\Psi_\alpha(\mathcal{R}_{j+1})| \\
&= \sum_{\alpha,\beta,\gamma} |\Psi_\beta(\mathcal{R}_{j+1})\rangle S^\dagger_{\beta\alpha} S_{\alpha\gamma}\langle\Psi_\gamma(\mathcal{R}_{j+1})|.
\end{aligned} \tag{B15}$$

If applied onto $|\Psi_\alpha(\mathcal{R}_0)\rangle$, then the Wilson loop operator transforms like $W[\mathcal{C}] \to SW[\mathcal{C}]S^\dagger$, so the eigenspectrum is invariant, too. In particular, we note that we do not have to find a real gauge of the eigenstates to compute the Wilson loop operator in contrast to when one attempts to compute the Euler curvature [10].

What if the Hamiltonian matrix $H$ given in some specific basis is not a real matrix? In that case, we can find a unitary transformation $R$ so that $H = R\tilde{H}R^\dagger$, where $\tilde{H}$ is a real matrix and $|\Psi\rangle = R|\tilde{\Psi}\rangle$, where $|\tilde{\Psi}\rangle$ are states that can have a real gauge. If $R$ and the new basis states do not depend on $\boldsymbol{\theta}$ then the overlaps in Eq. (B12) stay invariant. Because of this, the spectrum of the Wilson loop operator will always consist eigenvalues $+1$ and/or $-1$ and/or of TR partners $\{e^{i\phi}, e^{-i\phi}\}$.

If we parameterize the Hamiltonian with twisted boundary conditions, then $\hat{H}(\boldsymbol{\theta}) = \hat{H}(\theta_x + 2\pi, \theta_y) = \hat{H}(\theta_x, \theta_y + 2\pi)$ and the behavior of the Wilson loops will follow the above-described behavior. For different gauges, in particular the periodic gauge, this is a far more subtle matter. First of all, as explained in detail in Ref. [5], the Berry curvature depending on $\boldsymbol{\theta}$ is not gauge invariant with respect to gauge transformations in real space. Therefore, the Wilson loop operator is not gauge invariant under these transformations either, although the transformed Wilson loop operator will be in the same topological equivalence class, i.e. it will have the same winding. What is the advantage of either gauge (periodic or twisted boundary gauge)? In the twisted boundary gauge, we can straightforwardly compute Eq. (B11). The eigenstates in that gauge are translation invariant, however the translation operator depends on the twisted boundary phase. Therefore, eigenstates with different irreps $\boldsymbol{K}$ at different twist angles will not be orthogonal. This is important for the consideration of fractional Chern insulators, because the different eigenstates with different translation eigenvalues will, when computing the Wilson loop spectrum, hybridize

and the mathematical protection of the winding due to the Euler class does not seem to be protected. On the other hand, in the periodic gauge this should not happen and we should get a block-diagonal Wilson loop spectrum. However, in that case $\hat{H}(\boldsymbol{\theta})$ is not periodic in $2\pi$ anymore. We therefore need to find a gauge convention, as the Wilson loop operator spectrum is not going to be gauge invariant anymore. Using the large gauge transformation $\hat{U}_x$ (see Eq. (8) in the main text) we can transform $\hat{H}(2\pi, \theta_y) = \hat{U}_x \hat{H}(0, \theta_y) \hat{U}_x^\dagger$. We then get a gauge invariant Wilson loop operator, by defining the gauge convention

$$|\boldsymbol{\Psi}(2\pi)\rangle = \hat{U}_x |\boldsymbol{\Psi}(0)\rangle. \tag{B16}$$

This will still result in a real Wilson loop operator, provided that $\hat{U}_x$ commutes with $\hat{A}$.

**Appendix C: Real vector bundles**

In this section we deduce and review some topological properties associated to real wave functions defined on a torus (or more generally speaking "real vector bundles"). Readers who are not concerned with technical details may choose to skip this section.

**1. Real eigenstates and antiunitary symmetries**

We are given a family of Hermitian operators $\hat{H}(\boldsymbol{\theta})$ and an antiunitary operator $\hat{A} = C_{2z}\mathcal{T}$. This operator always squares to one[5] and on a 2D lattice commutes with the Hamiltonian for any $\boldsymbol{\theta}$. First, let us review some essential properties of antiunitary operators and of time reversal in particular. The action of any antiunitary operator $\hat{A}$ on some orthonormal basis states $\{|\alpha\rangle\}$ can be written as

$$\hat{A}c|\alpha\rangle = c^* U_{\alpha\beta}|\beta\rangle. \tag{C1}$$

Here, $U$ is some unitary base dependent matrix. This is often expressed as

$$\langle \alpha | \hat{A} | \beta \rangle = U_{\alpha\beta} K,$$
$$K|\alpha\rangle = |\alpha\rangle K, \tag{C2}$$

with complex conjugation $K$ (the second line expresses that possible scalars after the state $|\alpha\rangle$ have to be complex conjugated as well). The time reversal operator $\hat{\mathcal{T}}$ is an important antiunitary operator, which merely acts[6] as $\hat{\mathcal{T}} = K$ on spinless particles and as

$$\hat{\mathcal{T}} = e^{-i\pi \hat{S}_y} K \tag{C3}$$

on spinfull particles. Note that for spin-1/2 particles (with $\hat{S}_y = \hat{\sigma}_y/2$, $\hbar = 1$) we identify

$$\mathcal{T} = -i\hat{\sigma}_y K = (|\downarrow\rangle\langle\uparrow| - |\uparrow\rangle\langle\downarrow|)K, \tag{C4}$$

hence

$$\hat{\mathcal{T}} c_\uparrow \hat{\mathcal{T}}^{-1} = c_\downarrow, \qquad \hat{\mathcal{T}} c_\downarrow \hat{\mathcal{T}}^{-1} = -c_\uparrow$$
$$\hat{\mathcal{T}} c_\uparrow^\dagger \hat{\mathcal{T}}^{-1} = c_\downarrow^\dagger, \qquad \hat{\mathcal{T}} c_\downarrow^\dagger \hat{\mathcal{T}}^{-1} = -c_\uparrow^\dagger. \tag{C5}$$

Whenever $\hat{A}\hat{H}(\boldsymbol{\theta})\hat{A}^{-1} = \hat{H}(\boldsymbol{\theta})$ it is possible to find a representation $|\tilde{\alpha}\rangle$, where $\langle\tilde{\alpha}|\hat{H}|\tilde{\beta}\rangle = \tilde{H}_{\alpha\beta}$ is a real matrix [10]. Since $\hat{A}^2 = +1$ we readily identify $U = U^T$. It is then (according to the Autonne–Takagi factorization) possible to find a unitary matrix $V$ so that $U = VDV^T$, where $D$ is the matrix of eigenvalues of the form $e^{i\varphi_j}$. Now let us define $W = \sqrt{D^*}V^\dagger$ and $|\tilde{\alpha}\rangle = W^\dagger|\alpha\rangle$, then

$$\hat{A}|\tilde{\alpha}\rangle = VDV^T KV\left(\sqrt{D^*}\right)^*|\alpha\rangle = VD\sqrt{D^*}|\alpha\rangle K$$
$$= V\left(\sqrt{D^*}\right)^*|\alpha\rangle K = |\tilde{\alpha}\rangle K. \tag{C6}$$

We see that on this basis $\hat{A}$ simply acts as complex conjugation. Hence the Hamiltonian matrix $\tilde{H}$ has to be real, if $\hat{A}$ commutes with the Hamiltonian. We note that the transformation $W$ is independent of $\boldsymbol{\theta}$ and just corresponds to some unitary rotation of the Hamiltonian if $\hat{A}$ does not depend on $\boldsymbol{\theta}$.

**2. Euler and Stiefel-Whitney classes**

In topological band theory, characteristic classes describe a topological obstruction to define smooth, symmetric wave functions over the Brillouin zone [11]. Hence, they are related to some type of topological obstruction in the Wannier functions and manifest themselves in the Wilson loop spectrum [12, 13]. Characteristic classes associated to real vector bundles in 1D and 2D are the first and second Stiefel-Whitney classes and the Euler class [6, 11].

Let us start with Stiefel-Whitney (SW) classes. Characteristic classes are a subset of the cohomology classes of the base manifold of a fiber bundle. SW classes take their values in the Čech cohomology groups $H^r(M; \mathbb{Z}_2)$[7], but at least for the first Stiefel-Whitney class (SW1) we also have a representation in terms of de Rham cohomology, where the Berry phase modulo $2\pi$ is equivalent to

---

[5] This is obvious for many-body states with an even number of spin-1/2 particles transforming like an integer spin. For an odd number of particles note that $\hat{C}_{2z}$ squares to $-1$ due to the double-representation under which half-integer spin systems transform.

[6] For convenience we use this imprecise notation, even though $K$ is base dependent in contrast to $\hat{\mathcal{T}}$.

[7] Let $i_0, \ldots, i_r$ all have an open environment $U_{i_j}$, so that

$$U_{i_0} \cap \cdots \cap U_{i_r} \neq \emptyset.$$

Then we can define cochains $f(i_0, \ldots, i_r) \in \mathbb{Z}_2$ (where the function is invariant under arbitrary permutations of its arguments) and a suitable coboundary operator and with it cohomology groups in the usual sense. The transition functions of a fiber bundle then serve as cochains and yield a sense of orientation.

SW1[8] [12]. The SW1 is equivalent to the orientation of the fiber bundle, whereas the second Stiefel-Whitney class (SW2) tells us, whether we can find a "spin structure" (or "pin structure", if the bundle is not orientable) for the fiber bundle[9] [14]. Euler numbers on the other hand can be expressed in terms of de Rahm cohomology, so we have an associated curvature form. In general they are defined as the Pfaffian over the exterior derivative of the connection form $A = A_i dk_i$, so $\text{Pf}(dA)$. For a nontrivial Euler class, we need an even-dimensional fiber which has the same dimension as the base manifold[10]. Furthermore, it can be shown that SW2 is related to the Euler class by being nothing but the parity $(-1)^{e_2}$ of the Euler number $e_2$ in 2D. Without further symmetries, all topological invariants are described by Euler and SW classes [15]. In the present context however, the Euler number is of special significance as it can directly be linked to transport.

### 3. SW1 and orientability

States parameterized by $\boldsymbol{\theta}$ are special insofar as their base manifold has a non-trivial fundamental group. This means that there are noncontractible loops along which the fiber bundle may not be oriented. In addition, crystalline symmetries may indicate topological phases, as they do in band topology. As a "warm-up" we show[11] that the Berry phase (modulo $2\pi$) yields SW1 [12]. Let $\{|\Psi_n(\boldsymbol{\theta})\rangle\}$ be (possibly) complex eigenstates of a real hermitian Hamiltonian matrix $H(\boldsymbol{\theta})$. If we impose no symmetry-related restriction on these states, it is always possible to pick a smooth gauge for them[12]. Let us assume they are smooth on some noncontractible loop $\mathcal{C}$. We can make them real by a suitable gauge transformation $g(\boldsymbol{\theta})$

$$|\Psi_n^{\text{real}}(\boldsymbol{\theta})\rangle = |\Psi_m(\boldsymbol{\theta})\rangle g_{mn}(\boldsymbol{\theta}) \tag{C7}$$

Since the Berry connection is a hermitian matrix[13], its diagonal elements $\langle\Psi_m^{\text{real}}|i\boldsymbol{\nabla}_{\boldsymbol{\theta}}|\Psi_m^{\text{real}}\rangle$ in the real gauge vanish for smooth wave functions. Let us say, we have an open contour $\tilde{\mathcal{C}}$, which is equal to the non-contractable loop $\mathcal{C}$ with one point removed. The gauge transformation $g$ can be chosen so that $\{|\Psi_n^{\text{real}}(\boldsymbol{\theta})\rangle\}$ are smooth on $\tilde{\mathcal{C}}$ [17]. Then clearly

$$0 = \int_{\tilde{\mathcal{C}}} d\boldsymbol{\theta} \text{Tr} \boldsymbol{A}^{real}. \tag{C8}$$

We further identify

$$\begin{aligned}A_{mn}^{real} &= \langle\Psi_m^{\text{real}}(\boldsymbol{\theta})|i\boldsymbol{\nabla}_{\boldsymbol{\theta}}|\Psi_n^{\text{real}}(\boldsymbol{\theta})\rangle \\ &= g_{\alpha,m}^*(\boldsymbol{k})\langle\Psi_\alpha(\boldsymbol{\theta})|i\boldsymbol{\nabla}_{\boldsymbol{k}}|\Psi_\beta(\boldsymbol{\theta})\rangle g_{\beta n}(\boldsymbol{k}) \\ &= (g^\dagger i\boldsymbol{\nabla}_{\boldsymbol{k}} g)_{mn} + (g^\dagger A g)_{mn}\end{aligned} \tag{C9}$$

Let $g = S^\dagger D S$, $D$ is diagonal with eigenvalues $e^{i\gamma_{j,\boldsymbol{\theta}}}$; then

$$\begin{aligned}\text{Tr}(g^\dagger \boldsymbol{\nabla}_{\boldsymbol{\theta}} g) &= \text{Tr}(D^\dagger \boldsymbol{\nabla}_{\boldsymbol{\theta}} D) = i\sum_j \boldsymbol{\nabla}_{\boldsymbol{\theta}} \gamma_{j,\boldsymbol{\theta}} \\ &= i\boldsymbol{\nabla}_{\boldsymbol{\theta}}\text{Arg}(\det(g)) = \boldsymbol{\nabla}_{\boldsymbol{\theta}} \ln(\det(g)).\end{aligned} \tag{C10}$$

This leads to

$$\begin{aligned}0 = \int_{\tilde{\mathcal{C}}} d\boldsymbol{\theta} \text{Tr} \boldsymbol{A}^{real} &= \int_{\tilde{\mathcal{C}}} d\boldsymbol{\theta}(\text{Tr}\boldsymbol{A} + i\boldsymbol{\nabla}_{\boldsymbol{\theta}}\ln(\det(g))) \\ \Leftrightarrow \quad \frac{\det(g(\mathcal{C}_2))}{\det(g(\mathcal{C}_1))} &= \exp\left(i\int_{\tilde{\mathcal{C}}} d\boldsymbol{\theta}\text{Tr}\boldsymbol{A}\right),\end{aligned} \tag{C11}$$

where $\mathcal{C}_1, \mathcal{C}_2$ are the start- and endpoints of $\tilde{\mathcal{C}}$. Since $|\Psi_n(\mathcal{C}_1)\rangle = |\Psi_n(\mathcal{C}_2)\rangle$ we readily find

$$|\Psi_m^{\text{real}}(\mathcal{C}_1)\rangle \left(g^{-1}(\mathcal{C}_1)g(\mathcal{C}_2)\right)_{mn} = |\Psi_n^{real}(\mathcal{C}_2)\rangle, \tag{C12}$$

so the Berry phase is equivalent to the determinant of the transition function $g^{-1}(\mathcal{C}_1)g(\mathcal{C}_2)$ and hence tells us, whether the bundle is oriented or not.

### 4. Orientation of the AQHE bundle

To determine, whether the "AQHE bundle" — the fiber bundle obtained by two Chern insulating states related by $\hat{A} = \hat{C}_2\hat{\mathcal{T}}$ — is oriented or not, we compute the Berry phase in a smooth gauge. Such a gauge is provided by

---

[8] For the SW1, the integration over a non-contractable loop of the BZ is equivalent to the Berry Phase modulo $2\pi$, where $\pi$ denotes the nontrivial element.

[9] The consideration here is, whether it is possible to associate to the real bundle with structure group $SO(N)$ another bundle with structure group $Spin(N)$ (for non-orientable bundles analogously) that projects onto the original bundle, so that we can define transition functions over the entire manifold, in particular they must satisfy $\tilde{t}_{ij}\tilde{t}_{jk}\tilde{t}_{ki} = id$, with transition functions $\tilde{t}$ and the identity map $id$.

[10] In 2D this is easy to see, as the transition function between two patches is an element (assuming the bundle is oriented, which it has to be for a non-trivial Euler number) of $SO(N)$. The fundamental group of $SO(N)$ is $\mathbb{Z}_2$ for $N > 2$, but equal to $\mathbb{Z}$ for $N = 2$, hence we can get integer invariants in that case. For complex fiber bundles, note that the fundamental group of $U(N)$ is the same as $U(1)$ which is $\mathbb{Z}$, so one can always get integer Chern numbers.

[11] In fact this holds only if the SW1 of each one of the basis states of the Hamiltonian matrix is trivial. For a nontrivial basis neither Berry phase nor Wilson loop operator can detect a topologically nontrivial phase.

[12] This results from the fact that a Bloch bundle in 2D is trivial, if its Chern number vanishes [16]. Even for a nontrivial Chern number, states can still be defined smoothly on closed loops as long as the spectral projector on the loop is smooth [17].

[13] Note that

$$i\boldsymbol{\nabla}_{\boldsymbol{\theta}}\langle\Psi_n|\Psi_m\rangle = 0 = -(\langle\Psi_m|i\boldsymbol{\nabla}_{\boldsymbol{\theta}}|\Psi_n\rangle)^* + \langle\Psi_n|i\boldsymbol{\nabla}_{\boldsymbol{\theta}}|\Psi_m\rangle,$$

hence

$$(A_{mn})^* = A_{nm}.$$

the basis $|C = \pm 1\rangle$, provided no singularities are on the considered path [17]. We need to evaluate

$$\int_{\mathcal{C}} d\boldsymbol{\theta} Tr \boldsymbol{A}$$
$$= \int_{\mathcal{C}} d\boldsymbol{\theta} \langle C = 1|i\boldsymbol{\nabla_\theta}|C = 1\rangle + \langle C = -1|i\boldsymbol{\nabla_\theta}|C = -1\rangle. \quad (C13)$$

We note that

$$\boldsymbol{\nabla_\theta}\langle C = -1|C = -1\rangle = 0$$
$$\Rightarrow \langle C = -1|\boldsymbol{\nabla_\theta}|C = -1\rangle = -\langle C = -1|\boldsymbol{\nabla_\theta}|C = -1\rangle^*. \quad (C14)$$

Furthermore, assuming $\hat{A} = \hat{C}_2\hat{\mathcal{T}}$ does not explicitly depend on $\boldsymbol{\theta}$

$$\langle C = -1|\boldsymbol{\nabla_\theta}|C = -1\rangle = \langle \hat{A}C = 1|\boldsymbol{\nabla_\theta}\hat{A}C = 1\rangle$$
$$= \langle \hat{A}C = 1|\hat{A}\boldsymbol{\nabla_\theta}C = 1\rangle = \langle C = 1|\boldsymbol{\nabla_\theta}C = 1\rangle^*$$
$$= -\langle C = 1|\boldsymbol{\nabla_\theta}C = 1\rangle. \quad (C15)$$

We see that the Berry phase vanishes, so the SW1 is trivial.

### 5. Symmetry indicators

We finally comment on crystalline symmetries, in particular rotation symmetry. Other crystalline symmetries, such as mirror symmetry, however may also indicate Euler topology by protecting linear crossings [18].

How are rotation eigenvalues related to Euler numbers? Let us consider the case of two quasidegenerate states describing two Chern insulating partners under $\hat{A} = \hat{\mathcal{T}}\hat{C}_2$. These states will then also be eigenstates of the rotation operator as can be shown the following way. First, consider complex eigenvalues with non-zero imaginary part. It has to be possible to construct two ground states that are eigenstates of $\hat{C}_n$, $|\gamma\rangle$ and $|\gamma^*\rangle$, that are partners under $\hat{A}$, as

$$\hat{C}_n|\gamma\rangle = e^{i\gamma}|\gamma\rangle, \qquad \hat{C}_n(\hat{A}|\gamma\rangle) = e^{-i\gamma}(\hat{A}|\gamma\rangle), \quad (C16)$$

because $\hat{\mathcal{T}}C_2$ commutes with $C_n$. Note that $|\gamma\rangle \neq |\gamma^*\rangle$ if $e^{i\gamma} \neq e^{-i\gamma}$. We can also construct two Chern insulating states $|+C\rangle$ and $|-C\rangle$ that are partners under $\hat{A}$. Now let these two states be superpositions of the eigenstates of the rotation operator:

$$|+C\rangle = c_1|\gamma\rangle + c_2|\gamma^*\rangle$$
$$|-C\rangle = c_1'|\gamma\rangle + c_2'|\gamma^*\rangle. \quad (C17)$$

Because $\hat{A}|+C\rangle = |-C\rangle$ we identify $c_1 = c_2'^*$ and $c_2 = c_1'^*$. As $\langle +C| - C\rangle = 0$ we find $c_1 c_2 = 0$, so these states have to be eigenstates of the rotation operator. Why do the Chern insulating states have to be eigenstates of symmetry operators, if the eigenvalues are real? It is clear that they are eigenstates, if both eigenvalues are $+$ or $-$, so we only need to consider $+-$. Due to the fact that the "AQHE bundle" is orientable (see Sec. C 4), this implies that at each high-symmetry point the eigenvalues are $+-$, hence indicating spontaneous symmetry breaking of rotation symmetry and not an AQHE, see Sec. E.

If the AQHE bundle is deformable into two Chern insulators with opposite chirality in the above-described way, we can infer constraints on the Euler number using rotation symmetry. Let us assume $|+C\rangle$ transforms according to the irreps $\{\rho_{\Gamma_i}^{C_n}\}$ at the high-symmetry points $\{\Gamma_i\}$ of the rotation group $C_n$ and hence $|-C\rangle$ according to $\{(\rho_{\Gamma_i}^{C_n})^*\}$. Then we can relate the Chern number $C$ of $|+C\rangle$ to the rotation symmetry eigenvalues according to [19]

$$\begin{aligned} C_2 &: e^{-2\pi i C/2} = \rho_\Gamma^{C_2} \rho_M^{C_2} \rho_X^{C_2} \rho_Y^{C_2} \\ C_3 &: e^{-2\pi i C/3} = (-1)^{2SN_e} \rho_\Gamma^{C_3} \rho_K^{C_3} \rho_{K'}^{C_3} \\ C_4 &: e^{-2\pi i C/4} = (-1)^{2SN_e} \rho_\Gamma^{C_4} \rho_M^{C_4} \rho_X^{C_2} \\ C_6 &: e^{-2\pi i C/6} = (-1)^{2SN_e} \rho_\Gamma^{C_6} \rho_K^{C_3} \rho_M^{C_2}, \end{aligned} \quad (C18)$$

where $N_e$ is the number of particles and $S$ their spin. We have seen in the main paper that a Chern number $C$ implies an Euler number of the same value. By considering all different permutations of $\rho_{\Gamma_i}^{C_n}$ with their complex conjugated value $(\rho_{\Gamma_i}^{C_n})^*$, as we cannot a priori say, which irrep belongs to $|+C\rangle$ or $|-C\rangle$ we can go through all different combinations of possible Chern numbers and hence Euler numbers. The relation between rotation eigenvalues and Euler numbers are represented in Tab. I, II and III. For $C_2$ a nontrivial Euler number of 1 mod 2 is indicated by the parity of the number of (-1) eigenvalue pairs [12].

| $\Gamma$ | $K$ | $K'$ | Euler number |
|---|---|---|---|
| 1 | 1 | 1 | 0 mod 3 |
| 1 | 1 | $e^{\pm 2\pi i/3}$ | 1, 2 mod 3 |
| 1 | $e^{\pm 2\pi i/3}$ | 1 | 1, 2 mod 3 |
| $e^{\pm 2\pi i/3}$ | 1 | 1 | 1, 2 mod 3 |
| 1 | $e^{\pm 2\pi i/3}$ | $e^{\pm 2\pi i/3}$ | no constraint |
| $e^{\pm 2\pi i/3}$ | 1 | $e^{\pm 2\pi i/3}$ | no constraint |
| $e^{\pm 2\pi i/3}$ | $e^{\pm 2\pi i/3}$ | 1 | no constraint |
| $e^{\pm 2\pi i/3}$ | $e^{\pm 2\pi i/3}$ | $e^{\pm 2\pi i/3}$ | no constraint |

Table I. Symmetry indicators for $C_3$ rotation symmetry based on rotation eigenvalues at high-symmetry points.

### 6. Relation between Wilson loop flows and topological invariants

We close this appendix by addressing whether real topological invariants, in particular the SW2, can be recovered from Wilson loop flows, which describe the winding and symmetry protected crossings of the phases of the Wilson loop spectra. We closely follow [12]. Consider a real vector bundle defined on a sphere, so that the structure

| $\Gamma$ | $M$ | $X$ | Euler number |
|---|---|---|---|
| 1 | 1 | 1 | 0 mod 4 |
| $-1$ | $-1$ | 1 | 0 mod 4 |
| $-1$ | 1 | $-1$ | 0 mod 4 |
| 1 | $-1$ | $-1$ | 0 mod 4 |
| $-1$ | 1 | 1 | 2 mod 4 |
| 1 | $-1$ | 1 | 2 mod 4 |
| 1 | 1 | $-1$ | 2 mod 4 |
| $-1$ | $-1$ | $-1$ | 2 mod 4 |
| $\pm i$ | $\pm 1$ | $\pm 1$ | odd |
| $\pm 1$ | $\pm i$ | $\pm 1$ | odd |
| $\pm i$ | $\pm i$ | $\pm 1$ | even |

Table II. Symmetry indicators for $C_4$ rotation symmetry based on rotation eigenvalues at high-symmetry points.

| $\Gamma$ | $K$ | $M$ | Euler number |
|---|---|---|---|
| 1 | 1 | 1 | 0 mod 6 |
| $-1$ | 1 | $-1$ | 0 mod 6 |
| $-1$ | 1 | 1 | 3 mod 6 |
| 1 | 1 | $-1$ | 3 mod 6 |
| $e^{\pm 2\pi i/3}$ | 1 | 1 | 2, 4 mod 6 |
| $e^{\pm 2\pi i/6}$ | 1 | $-1$ | 2, 4 mod 6 |
| 1 | $e^{\pm 2\pi i/3}$ | 1 | 2, 4 mod 6 |
| $-1$ | $e^{\pm 2\pi i/3}$ | $-1$ | 2, 4 mod 6 |
| $e^{\pm 2\pi i/6}$ | 1 | 1 | 1, 5 mod 6 |
| $e^{\pm 2\pi i/3}$ | 1 | $-1$ | 1, 5 mod 6 |
| $-1$ | $e^{\pm 2\pi i/3}$ | 1 | 1, 5 mod 6 |
| 1 | $e^{\pm 2\pi i/3}$ | $-1$ | 1, 5 mod 6 |
| $e^{\pm 2\pi i/6}$ | $e^{\pm 2\pi i/3}$ | 1 | odd |
| $e^{\pm 2\pi i/3}$ | $e^{\pm 2\pi i/3}$ | $-1$ | odd |
| $e^{\pm 2\pi i/3}$ | $e^{\pm 2\pi i/3}$ | 1 | even |
| $e^{\pm 2\pi i/6}$ | $e^{\pm 2\pi i/3}$ | $-1$ | even |

Table III. Symmetry indicators for $C_6$ rotation symmetry based on rotation eigenvalues at high-symmetry points.

group[14] is $SO(N)$. SW2 then describes, whether it is possible to define a fiber bundle with structure group $Spin(N)$, that acts as a double cover of the original fiber bundle. In particular, one has to be able to define transition functions $\tilde{t}^{A_i A_j}$ between any two overlapping patches $x \in A_i \cap A_j$, so that

$$\tilde{t}^{AB}(x)\tilde{t}^{BC}(x)\tilde{t}^{CA}(x) = id \quad (C19)$$

with the identity map $id$. We can cover a sphere, parameterized by the two angles $\varphi \in [0, 2\pi]$ and $\vartheta \in [0, \pi]$, by three patches $A, B, C$, so that $A$ and $B$ overlap at $\varphi = 0$, $A$ and $C$ overlap at $\varphi = \pi/2$, $B$ and $C$ overlap at $\varphi = \pi$ and all three patches overlap at $\vartheta = 0, \pi$. Then we can relate

$$(\mathcal{R}(2\pi))^{w_2} = \tilde{f}^{ABC}(0)\tilde{f}^{ABC}(\pi), \quad (C20)$$

---

[14] SW1 has to be trivial, since $\pi_1(S^2) = 0$.

where $\mathcal{R}(2\pi)$ denotes a $2\pi$ rotation in $Spin(N)$, $w_2$ is the second Stiefel-Whitney class and

$$\tilde{f}^{ABC}(\vartheta) = \tilde{t}^{AB}(\varphi^{AB}, \vartheta)\tilde{t}^{BC}(\varphi^{BC}, \vartheta)\tilde{t}^{CA}(\varphi^{CA}, \vartheta)$$
$$\varphi^{AB} = 0, \quad \varphi^{BC} = \pi, \quad \varphi^{CA} = \pi/2. \quad (C21)$$

Since in $SO(N)$ a $2\pi$ rotation is just the identity, the projection of $\tilde{f}^{ABC}$ in the original bundle is defined as

$$f^{ABC}(\vartheta) = t^{AB}(\varphi^{AB}, \vartheta)t^{BC}(\varphi^{BC}, \vartheta)t^{CA}(\varphi^{CA}, \vartheta), \quad (C22)$$

where $t^{A_i A_j}$ are transition functions $\in SO(N)$ and $f^{ABC}(0) = f^{ABC}(\pi) = id$. If $\tilde{f}^{ABC}$ is connecting $id$ with $\mathcal{R}(2\pi)$ by varying $\vartheta$ from 0 to $\pi$, then this will be indicated by an odd winding number of the flow of $f^{ABC}$. This flow will be protected by an odd number of pairs of $(e^{i\pi}, e^{-i\pi})$ eigenvalues of $f^{ABC}$ at various $\vartheta$.

How can this be related to Wilson loop operators? Using the parallel transport gauge

$$|\boldsymbol{\Psi}_{\text{parallel}}(\phi, \vartheta)\rangle = |\boldsymbol{\Psi}(\phi, \vartheta)\rangle \hat{P}_{\text{path ordering}} e^{-i \int_0^\phi d\phi' A_\phi(\phi', \vartheta)} \quad (C23)$$

it is straightforward to show that the Wilson loop operator is then given by

$$\begin{aligned} W(\vartheta) &= \langle \boldsymbol{\Psi}^A_{\text{parallel}}(0, \vartheta) | \boldsymbol{\Psi}^B_{\text{parallel}}(2\pi, \vartheta)\rangle \\ &\cdot \langle \boldsymbol{\Psi}^B_{\text{parallel}}(\pi, \vartheta) | \boldsymbol{\Psi}^C_{\text{parallel}}(\pi, \vartheta)\rangle \\ &\cdot \langle \boldsymbol{\Psi}^C_{\text{parallel}}(\pi/2, \vartheta) | \boldsymbol{\Psi}^A_{\text{parallel}}(\pi/2, \vartheta)\rangle \\ &= t_p^{AB}(0, \vartheta) t_p^{BC}(\pi, \vartheta) t_p^{CA}(\pi/2, \vartheta), \quad (C24) \end{aligned}$$

with transition functions $t_p^{A_i A_j}$ in parallel transport gauge. Since topological invariants don't change under smooth deformations, we can deform $W(\vartheta)$ to $f^{ABC}(\vartheta)$ continuously. A similar relation can be identified for the Wilson loop operator $W(\theta_y)$ on a torus. In contrast to the sphere, there is no $\theta_y$, where $W(\theta_y)$ must be identical to the identity map. This implies that even though a nontrivial SW2 is still indicated by pairs of eigenvalues $(e^{i\pi}, e^{-i\pi})$ of $W(\theta_y)$, there does not have to be a nonzero winding number associated to it.

## Appendix D: Hamiltonian matrices and symmetry operators

### 1. Hamiltonian matrix elements

To project states to the lower band of Eq. (9) of the main text, we have to perform a Fourier transformation. Our unit cell consists of two neighboring sites in $x$ direction. We define two sublattices $A$ and $B$ so that $c^\dagger_{A,(R_x,R_y),\sigma} = c^\dagger_{x,y,\sigma}$ for $(x, y) = (2R_x - 1, R_y)$ and $c^\dagger_{B,(R_x,R_y),\sigma} = c^\dagger_{x,y,\sigma}$ for $(x, y) = (2R_x, R_y)$. The Fourier transformation of these operators is given by

$$c^\dagger_{\alpha,(R_x,R_y),\sigma} = \frac{1}{\sqrt{N}} \sum_{\boldsymbol{k}} e^{-i2k_x R_x} e^{-ik_y R_y} c^\dagger_{\alpha,\boldsymbol{k},\sigma}, \quad (D1)$$

where $N = N_x N_y$ and[15]

$$k_x = \frac{\pi}{N_x} j_x, \qquad k_y = \frac{2\pi}{N_y} j_y,$$
$$j_x = 0, \ldots, N_x - 1, \qquad j_y = 0, \ldots, N_y - 1. \qquad (D2)$$

For the periodic gauge we multiply to the hoppings in $x$ and $y$ direction the phase factors $e^{-i\theta_x/2N_x}$ and $e^{-i\theta_y/N_y}$, respectively. We note that this *does not* correspond to just replacing $k_{x/y}$ with $k_{x/y} + \theta_{x/y}/N_{x/y}$, because of the enlarged unit cell. In reciprocal space Eq. (9) becomes

$$\hat{H}_{\text{periodic}}(\boldsymbol{\theta})$$
$$= \sum_{\boldsymbol{k}} \left( c^\dagger_{A,\boldsymbol{k},\uparrow}, c^\dagger_{B,\boldsymbol{k},\uparrow}, c^\dagger_{A,\boldsymbol{k},\downarrow}, c^\dagger_{B,\boldsymbol{k},\downarrow} \right) h(\boldsymbol{k},\boldsymbol{\theta}) \begin{pmatrix} c_{A,\boldsymbol{k},\uparrow} \\ c_{B,\boldsymbol{k},\uparrow} \\ c_{A,\boldsymbol{k},\downarrow} \\ c_{B,\boldsymbol{k},\downarrow} \end{pmatrix}, \qquad (D3)$$

where

$$\begin{aligned}
h_{11} &= \cos(2\pi\gamma)(2t\cos(k_y + \theta_y/N_y)) \\
h_{12} &= \cos(2\pi\gamma)\left[-t\left(e^{-2ik_x - i\frac{\theta_x}{2N_x}} + e^{i\frac{\theta_x}{2N_x}}\right)\right] \\
&\quad + 2t'\sin(k_y + \frac{\theta_y}{N_y})\left(e^{-2ik_x - i\frac{\theta_x}{2N_x}} - e^{i\frac{\theta_x}{2N_x}}\right) \\
h_{13} &= (-i\sin(2\pi\gamma))\left(-2t\sin(k_y + \frac{\theta_y}{N_y})\right) \\
h_{14} &= (-i\sin(2\pi\gamma))\left[-t\left(e^{-2ik_x - i\frac{\theta_x}{2N_x}} + e^{i\frac{\theta_x}{2N_x}}\right)\right] \\
h_{22} &= \cos(2\pi\gamma)\left(-2t\cos(k_y + \frac{\theta_y}{N_y})\right) \\
h_{23} &= (-i\sin(2\pi\gamma))\left[-t\left(e^{-i\frac{\theta_x}{2N_x}} - e^{2ik_x + i\frac{\theta_x}{2N_x}}\right)\right] \\
h_{24} &= (-i\sin(2\pi\gamma))\left(2t\sin(k_y + \frac{\theta_y}{N_y})\right) \\
h_{33} &= \cos(2\pi\gamma)\left(2t\cos(k_y + \frac{\theta_y}{N_y})\right) \\
h_{34} &= \cos(2\pi\gamma)\left[-t\left(e^{-i\frac{\theta_x}{2N_x}}e^{-2ik_x} + e^{i\frac{\theta_x}{2N_x}}\right)\right] \\
&\quad + 2t'\sin(k_y + \frac{\theta_y}{N_y})\left(-e^{-i\frac{\theta_x}{2N_x}}e^{-2ik_x} + e^{i\frac{\theta_x}{2N_x}}\right) \\
h_{44} &= \cos(2\pi\gamma)\left(-2t\cos(k_y + \frac{\theta_y}{N_y})\right).
\end{aligned} \qquad (D4)$$

How do eigenstates look like? We define

$$\epsilon(\boldsymbol{k},\boldsymbol{\theta}) = U_{\boldsymbol{k},\boldsymbol{\theta}} h(\boldsymbol{k},\boldsymbol{\theta}) U^\dagger_{\boldsymbol{k},\boldsymbol{\theta}} \qquad (D5)$$

and diagonalize the Hamiltonian

$$\hat{H}_{\text{periodic}}(\boldsymbol{\theta}) = \sum_{\boldsymbol{k}} (U_{\boldsymbol{k},\boldsymbol{\theta}} \boldsymbol{c}_{\boldsymbol{k}})^\dagger \epsilon(\boldsymbol{k},\boldsymbol{\theta}) U_{\boldsymbol{k},\boldsymbol{\theta}} \boldsymbol{c}_{\boldsymbol{k}}$$
$$= \sum_{\boldsymbol{k}} \sum_\alpha \tilde{c}^\dagger_{\alpha,\boldsymbol{k},\boldsymbol{\theta}} \epsilon_\alpha(\boldsymbol{k},\boldsymbol{\theta}) \tilde{c}_{\alpha,\boldsymbol{k},\boldsymbol{\theta}}. \qquad (D6)$$

Eigenstates are then

$$\prod_{(\alpha,\boldsymbol{k})} \tilde{c}^\dagger_{\alpha,\boldsymbol{k},\boldsymbol{\theta}} |0\rangle. \qquad (D7)$$

To project everything onto the lowest band, we need to work in $\boldsymbol{k}$ space and in the eigenbasis. We have two different types of interaction: Hubbard interaction, which is responsible for the spin polarization and Coulomb interaction for stabilizing a fractional Chern-insulating phase. In real space we have

$$\begin{aligned}
\hat{H}_{\text{Hubbard}} &= U \sum_{\boldsymbol{R}} \sum_{\gamma = A,B} \hat{n}_{\gamma,\boldsymbol{R},\uparrow} \hat{n}_{\gamma,\boldsymbol{R},\downarrow} \\
&= -U \sum_{\boldsymbol{R}} \sum_{\gamma = A,B} c^\dagger_{\gamma,\boldsymbol{R},\uparrow} c^\dagger_{\gamma,\boldsymbol{R},\downarrow} c_{\gamma,\boldsymbol{R},\uparrow} c_{\gamma,\boldsymbol{R},\downarrow}
\end{aligned} \qquad (D8)$$

and

$$\begin{aligned}
&\hat{H}_{\text{Coulomb, NN}} \\
&= V \sum_{\boldsymbol{R}} \hat{n}_{A,\boldsymbol{R}} \hat{n}_{A,(R_x, R_y+1)} + \hat{n}_{A,\boldsymbol{R}} \hat{n}_{B,\boldsymbol{R}} \\
&\quad + \hat{n}_{B,\boldsymbol{R}} \hat{n}_{B,(R_x, R_y+1)} + \hat{n}_{B,\boldsymbol{R}} \hat{n}_{A,(R_x+1, R_y)} \\
&= -V \sum_{\boldsymbol{R}} \sum_{\sigma,\sigma'} \Big\{ c^\dagger_{A,\boldsymbol{R},\sigma} c^\dagger_{A,(R_x, R_y+1),\sigma'} c_{A,\boldsymbol{R},\sigma} c_{A,(R_x, R_y+1),\sigma'} \\
&\quad + c^\dagger_{A,\boldsymbol{R},\sigma} c^\dagger_{B,\boldsymbol{R},\sigma'} c_{A,\boldsymbol{R},\sigma} c_{B,\boldsymbol{R},\sigma'} \\
&\quad + c^\dagger_{B,\boldsymbol{R},\sigma} c^\dagger_{B,(R_x, R_y+1),\sigma'} c_{B,\boldsymbol{R},\sigma} c_{B,(R_x, R_y+1),\sigma'} \\
&\quad + c^\dagger_{A,(R_x+1, R_y),\sigma} c^\dagger_{B,\boldsymbol{R},\sigma'} c_{A,(R_x+1, R_y),\sigma} c_{B,\boldsymbol{R},\sigma'} \Big\}
\end{aligned} \qquad (D9)$$

Using the Fourier transformation in Eq. (D1) we rewrite these interaction terms as

$$\begin{aligned}
&\hat{H}_{\text{Hubbard}} \\
&= -U \sum_{\boldsymbol{k}_1, \boldsymbol{k}_2, \boldsymbol{k}_3, \boldsymbol{k}_4} \sum_{\gamma = A,B} c^\dagger_{\gamma,\boldsymbol{k}_1,\uparrow} c^\dagger_{\gamma,\boldsymbol{k}_2,\downarrow} c_{\gamma,\boldsymbol{k}_3,\uparrow} c_{\gamma,\boldsymbol{k}_4,\downarrow} \\
&\quad \cdot \frac{2}{N} \delta^{\text{mod } 2\pi}_{2(k^1_x + k^2_x - k^3_x - k^4_x),0} \delta^{\text{mod } 2\pi}_{(k^1_y + k^2_y - k^3_y - k^4_y),0}
\end{aligned} \qquad (D10)$$

and

$$\begin{aligned}
&\hat{H}_{\text{Coulomb, NN}} \\
&= -\frac{2V}{N} \sum_{\sigma,\sigma'} \sum_{\boldsymbol{k}_1, \boldsymbol{k}_2, \boldsymbol{k}_3, \boldsymbol{k}_4} \delta^{\text{mod } 2\pi}_{2(k^1_x + k^2_x - k^3_x - k^4_x),0} \delta^{\text{mod } 2\pi}_{(k^1_y + k^2_y - k^3_y - k^4_y),0} \\
&\quad \cdot \Big\{ e^{-i(k^2_y - k^4_y)} c^\dagger_{A,\boldsymbol{k}_1,\sigma} c^\dagger_{A,\boldsymbol{k}_2,\sigma'} c_{A,\boldsymbol{k}_3,\sigma} c_{A,\boldsymbol{k}_4,\sigma'} \\
&\quad + (1 + e^{-i2(k^2_x - k^3_x)}) c^\dagger_{A,\boldsymbol{k}_1,\sigma} c^\dagger_{B,\boldsymbol{k}_2,\sigma'} c_{A,\boldsymbol{k}_3,\sigma} c_{B,\boldsymbol{k}_4,\sigma'} \\
&\quad + e^{-i(k^2_y - k^4_y)} c^\dagger_{B,\boldsymbol{k}_1,\sigma} c^\dagger_{B,\boldsymbol{k}_2,\sigma'} c_{B,\boldsymbol{k}_3,\sigma} c_{B,\boldsymbol{k}_4,\sigma'} \Big\}
\end{aligned} \qquad (D11)$$

As

$$c_{i,\boldsymbol{k}} = (U^\dagger_{\boldsymbol{k},\boldsymbol{\theta}})_{ij} \tilde{c}_{j,\boldsymbol{k},\boldsymbol{\theta}}, \qquad (D12)$$

where $U_{(\ldots)}$ is a $4 \times 4$ matrix with basis $((A,\uparrow), (B,\uparrow), (A,\downarrow), (B,\downarrow))$ we identify

$\hat{H}_{\text{Hubbard}}$

$$= -\frac{2U}{N} \sum_{\boldsymbol{k}_1,\boldsymbol{k}_2,\boldsymbol{k}_3,\boldsymbol{k}_4} \sum_{j_1,j_2,j_3,j_4=1}^{2} \delta_{2(k_x^1+k_x^2-k_x^3-k_x^4),0}^{\text{mod } 2\pi} \delta_{(k_y^1+k_y^2-k_y^3-k_y^4),0}^{\text{mod } 2\pi} \left\{ (U_{\boldsymbol{k}_1,\boldsymbol{\theta}}^\dagger)_{1,j_1}^* (U_{\boldsymbol{k}_2,\boldsymbol{\theta}}^\dagger)_{3,j_2}^* (U_{\boldsymbol{k}_3,\boldsymbol{\theta}}^\dagger)_{1,j_3} (U_{\boldsymbol{k}_4,\boldsymbol{\theta}}^\dagger)_{3,j_4} \right.$$
$$\left. + (U_{\boldsymbol{k}_1,\boldsymbol{\theta}}^\dagger)_{2,j_1}^* (U_{\boldsymbol{k}_2,\boldsymbol{\theta}}^\dagger)_{4,j_2}^* (U_{\boldsymbol{k}_3,\boldsymbol{\theta}}^\dagger)_{2,j_3} (U_{\boldsymbol{k}_4,\boldsymbol{\theta}}^\dagger)_{4,j_4} \right\} \tilde{c}_{j_1,\boldsymbol{k}_1,\boldsymbol{\theta}}^\dagger \tilde{c}_{j_2,\boldsymbol{k}_2,\boldsymbol{\theta}}^\dagger \tilde{c}_{j_3,\boldsymbol{k}_3,\boldsymbol{\theta}} \tilde{c}_{j_4,\boldsymbol{k}_4,\boldsymbol{\theta}} \quad \text{(D13)}$$

and

$$\hat{H}_{\text{Coulomb, NN}} = -\frac{2V}{N} \sum_{\boldsymbol{k}_1,\boldsymbol{k}_2,\boldsymbol{k}_3,\boldsymbol{k}_4} \sum_{j_1,j_2,j_3,j_4=1}^{2} \delta_{2(k_x^1+k_x^2-k_x^3-k_x^4),0}^{\text{mod } 2\pi} \delta_{(k_y^1+k_y^2-k_y^3-k_y^4),0}^{\text{mod } 2\pi}$$
$$\left[ e^{-i(k_y^2-k_y^4)} \left( (U_{\boldsymbol{k}_1,\boldsymbol{\theta}}^\dagger)_{1,j_1}^* (U_{\boldsymbol{k}_3,\boldsymbol{\theta}}^\dagger)_{1,j_3} + (U_{\boldsymbol{k}_1,\boldsymbol{\theta}}^\dagger)_{3,j_1}^* (U_{\boldsymbol{k}_3,\boldsymbol{\theta}}^\dagger)_{3,j_3} \right) \right.$$
$$\left( (U_{\boldsymbol{k}_2\boldsymbol{\theta}}^\dagger)_{1,j_2}^* (U_{\boldsymbol{k}_4,\boldsymbol{\theta}}^\dagger)_{1,j_4} + (U_{\boldsymbol{k}_2,\boldsymbol{\theta}}^\dagger)_{3,j_2}^* (U_{\boldsymbol{k}_4,\boldsymbol{\theta}}^\dagger)_{3,j_4} \right)$$
$$+ (1 + e^{-i2(k_x^1-k_x^3)}) \left( (U_{\boldsymbol{k}_1,\boldsymbol{\theta}}^\dagger)_{1,j_1}^* (U_{\boldsymbol{k}_3,\boldsymbol{\theta}}^\dagger)_{1,j_3} + (U_{\boldsymbol{k}_1,\boldsymbol{\theta}}^\dagger)_{3,j_1}^* (U_{\boldsymbol{k}_3,\boldsymbol{\theta}}^\dagger)_{3,j_3} \right)$$
$$\left( (U_{\boldsymbol{k}_2\boldsymbol{\theta}}^\dagger)_{2,j_2}^* (U_{\boldsymbol{k}_4,\boldsymbol{\theta}}^\dagger)_{2,j_4} + (U_{\boldsymbol{k}_2,\boldsymbol{\theta}}^\dagger)_{4,j_2}^* (U_{\boldsymbol{k}_4,\boldsymbol{\theta}}^\dagger)_{4,j_4} \right)$$
$$e^{-i(k_y^2-k_y^4)} \left( (U_{\boldsymbol{k}_1,\boldsymbol{\theta}}^\dagger)_{2,j_1}^* (U_{\boldsymbol{k}_3,\boldsymbol{\theta}}^\dagger)_{2,j_3} + (U_{\boldsymbol{k}_1,\boldsymbol{\theta}}^\dagger)_{4,j_1}^* (U_{\boldsymbol{k}_3,\boldsymbol{\theta}}^\dagger)_{4,j_3} \right)$$
$$\left. \left( (U_{\boldsymbol{k}_2\boldsymbol{\theta}}^\dagger)_{2,j_2}^* (U_{\boldsymbol{k}_4,\boldsymbol{\theta}}^\dagger)_{2,j_4} + (U_{\boldsymbol{k}_2,\boldsymbol{\theta}}^\dagger)_{4,j_2}^* (U_{\boldsymbol{k}_4,\boldsymbol{\theta}}^\dagger)_{4,j_4} \right) \right] \tilde{c}_{j_1,\boldsymbol{k}_1,\boldsymbol{\theta}}^\dagger \tilde{c}_{j_2,\boldsymbol{k}_2,\boldsymbol{\theta}}^\dagger \tilde{c}_{j_3,\boldsymbol{k}_3,\boldsymbol{\theta}} \tilde{c}_{j_4,\boldsymbol{k}_4,\boldsymbol{\theta}} \quad \text{(D14)}$$

### 2. Rotation symmetry

Regarding symmetry, the operator $\hat{C}_4^+$ acts like[16]:

$$\hat{C}_4^+ c_{A,\boldsymbol{R},\sigma}^\dagger \left(\hat{C}_4^+\right)^\dagger$$
$$= \begin{cases} R_y \text{ even:} & -c_{A,(\frac{N_y+2-R_y}{2},-1+2R_x),\sigma}^\dagger e^{-i\frac{\pi}{4}\sigma} \\ R_y \text{ odd:} & c_{B,(\frac{N_y+1-R_y}{2},-1+2R_x),\sigma}^\dagger e^{-i\frac{\pi}{4}\sigma} \end{cases}$$
$$\hat{C}_4^+ c_{B,\boldsymbol{R},\sigma}^\dagger \left(\hat{C}_4^+\right)^\dagger$$
$$= \begin{cases} R_y \text{ even:} & c_{A,(\frac{N_y+2-R_y}{2},2R_x),\sigma}^\dagger e^{-i\frac{\pi}{4}\sigma} \\ R_y \text{ odd:} & c_{B,(\frac{N_y+1-R_y}{2},2R_x),\sigma}^\dagger e^{-i\frac{\pi}{4}\sigma} \end{cases}$$
(D15)

In reciprocal space, the action of $\hat{C}_4^\dagger$ is given by

$$\hat{C}_4^+ c_{A,\boldsymbol{k},\sigma}^\dagger \left(\hat{C}_4^+\right)^\dagger = \frac{e^{-i\frac{\pi}{4}\sigma}}{2} e^{ik_x}$$
$$\cdot \left( e^{ik_y} c_{B,(-k_y \text{ mod } \pi,k_x),\sigma}^\dagger - e^{ik_y} c_{B,(-k_y \text{ mod } \pi,k_x+\pi),\sigma}^\dagger \right.$$
$$\left. -e^{2ik_y} c_{A,(-k_y \text{ mod } \pi,k_x),\sigma}^\dagger + e^{2ik_y} c_{A,(-k_y \text{ mod } \pi,k_x+\pi),\sigma}^\dagger \right)$$
$$\hat{C}_4^+ c_{B,\boldsymbol{k},\sigma}^\dagger \left(\hat{C}_4^+\right)^\dagger = \frac{e^{-i\frac{\pi}{4}\sigma}}{2}$$
$$\cdot \left( e^{ik_y} c_{B,(-k_y \text{ mod } \pi,k_x),\sigma}^\dagger + e^{ik_y} c_{B,(-k_y \text{ mod } \pi,k_x+\pi),\sigma}^\dagger \right.$$
$$\left. +e^{2ik_y} c_{A,(-k_y \text{ mod } \pi,k_x),\sigma}^\dagger + e^{2ik_y} c_{A,(-k_y \text{ mod } \pi,k_x+\pi),\sigma}^\dagger \right).$$
(D16)

### 3. Large gauge transformation

Next, we also need to find a representation of the large gauge transformation for evaluating the Wilson loop operator. Using the $A$ and $B$ sublattice, Eq. (8) of the main text becomes

$$\hat{U}_x = \exp\left( \frac{-2\pi i}{N_x} \sum_{\boldsymbol{R}} \sum_{\gamma=A,B} \sum_{\sigma} (2R_x - 2 + \gamma) c_{\gamma,\boldsymbol{R},\sigma}^\dagger c_{\gamma,\boldsymbol{R},\sigma} \right),$$
(D17)

with $\gamma = 1$ for the $A$ sublattice and $\gamma = 2$ for the $B$ sublattice. Its action on the creation operators is given

---

[15] For quadratic clusters we have $N_x = N_y/2$. Due to the enlarged unit cell of the Hofstadter model $k_x \in [0,\pi]$.

[16] Note that this form guarantees $(\hat{C}_4^+)^4 = +1$: We get a -1 due to the fact that we have spin-1/2 particles and another -1 due to an Aharonov-Bohm phase, when rotating a particle around a plaquette.

by

$$\hat{U}_x \tilde{c}^\dagger_{i,\boldsymbol{k},\boldsymbol{\theta}} \hat{U}^\dagger_x = \sum_j \sum_\alpha (U_{\boldsymbol{k},\boldsymbol{\theta}})^*_{i,\alpha} M_\alpha$$
$$\cdot (U^\dagger_{(k_x - 2\pi/N_x, k_y),\boldsymbol{\theta}})^*_{\alpha,j} \tilde{c}^\dagger_{j,(k_x - 2\pi/N_x, k_y),\boldsymbol{\theta}} \quad \text{(D18)}$$

with

$$M_\alpha = e^{\frac{2\pi i}{N_x}}, \quad \text{if } \alpha = (A, \uparrow) \text{ or } \alpha = (A, \downarrow). \quad \text{(D19)}$$

**Appendix E: Spontaneous symmetry breaking**

We want to give some final comments on spontaneous symmetry breaking. Measuring the Hall conductivity of a TRI system on a torus geometry will automatically break TRI, as applying a small electric field in some direction implies a magnetic flux due to Lenz's law. Therefore, as mentioned in the main text, we treat $\hat{A} = \hat{\mathcal{T}}\hat{C}_2$ as a generalization of time reversal, as it remains a symmetry for all fluxes $\boldsymbol{\theta}$. In the following we argue why the physical ground state is some $\hat{A}$-broken state.

Spontaneous symmetry breaking does occur, if energetic splittings between the fully-symmetric ground states become negligible and thermal fluctuations don't let one ground state tunnel into another one. In the main text we have seen that even on a relatively small $4 \times 4$ cluster, finite-size splitting between the ground states is tiny ($< 10^{-4} t$), leading to no significant energy penalty for the symmetry-broken state; in addition, for an odd number of particles, transforming like a half-integer wave function, $\hat{I}\hat{\mathcal{T}}$ would lead to a Kramers' degeneracy at all $\boldsymbol{\theta}$. The ground states are furthermore protected against local (thermal) fluctuations due to their ferromagnetic off-diagonal long-range order[17].

Why is no crystalline symmetry such as rotation symmetry spontaneously broken in contrast to $\hat{A}$? For locally ordered phases, adiabatically connected to atomic limits, such as charge-density-wave phases, one would expect to find different rotation irreps at all high-symmetry points, which is not what we found in the main text[18]. In addition, we have shown earlier that nontrivial Euler topology implies symmetry-protected crossings between the ground states for some $\boldsymbol{\theta}$. In the thermodynamic limit, it is known that the dependence of observables such as energy on $\boldsymbol{\theta}$ vanishes exponentially [23], ruling out splittings between the ground states induced by symmetry-breaking terms, as long as $\hat{A} = \hat{\mathcal{T}}\hat{C}_2$ is conserved.

---

[17] Such a protection even holds for certain models of spinless particles in the form of loop current correlation functions [20]. Even in the absence of any off-diagonal long-range order, a type of "nonlocal" order can be observed in quantum Hall systems [21, 22].
[18] We identified different irreps at $\boldsymbol{\theta} = (0, 0)$, but the same at $\boldsymbol{\theta} = (\pi, \pi)$.



\end{document}